\newcommand{\be}{\begin{equation}}
\newcommand{\ee}{\end{equation}}

\documentclass[journal,10pt]{IEEEtran}

\usepackage{booktabs}
\usepackage{psfrag}
\ifCLASSINFOpdf
   \usepackage[pdftex]{graphicx}

\else
  
   \usepackage[dvips]{graphicx}

\fi
\usepackage[cmex10]{amsmath}
\usepackage{amsthm,epstopdf,nicefrac,pgfplots}

  \pgfplotsset{compat=newest} 
  \pgfplotsset{plot coordinates/math parser=false}

\usepackage{rotating}

\usepackage{mathtools}

\usepackage{gensymb}

\renewcommand\appendix{\par
  \setcounter{section}{0}
  \setcounter{subsection}{0}
  \setcounter{figure}{0}
  \setcounter{table}{0}
  \renewcommand\thesection{Appendix \Alph{section}}
  \renewcommand\thefigure{\Alph{section}\arabic{figure}}
  \renewcommand\thetable{\Alph{section}\arabic{table}}
}

\usepackage{color}

\twocolumn

  \newlength\fheight 
    \newlength\fwidth 
\usepackage{balance}
\usepackage{algorithm}
\usepackage{algpseudocode}

\usepackage[bookmarks=false]{hyperref}
\begin{document}
%
\title{\color{black} Dynamic Signal Measurements\\ based on Quantized Data}
%
%
%
%


\newcommand\copyrighttext{%
  \footnotesize \textcopyright 2017 IEEE.  Personal use of this material is permitted. Permission from IEEE must be obtained for all other uses, in any current or future media, including reprinting/republishing this material for advertising or promotional purposes, creating new collective works, for resale or redistribution to servers or lists, or reuse of any copyrighted component of this work in other works. 
 DOI: \href{https://doi.org/10.1109/TIM.2016.2627298}{10.1109/TIM.2016.2627298}
 }
\newcommand\copyrightnotice{%
\begin{tikzpicture}[remember picture,overlay]
\node[anchor=south,yshift=10pt] at (current page.south) {\fbox{\parbox{\dimexpr\textwidth-\fboxsep-\fboxrule\relax}{\copyrighttext}}};
\end{tikzpicture}%
}

\author{P.~Carbone,~\IEEEmembership{Fellow Member,~IEEE}\thanks{P. Carbone and Antonio Moschitta are with the University of Perugia - Engineering Department, via G. Duranti, 93 - 06125 Perugia Italy,}
and~J.~Schoukens,~\IEEEmembership{Fellow Member,~IEEE}\thanks{J. Schoukens is with the Vrije Universiteit Brussel, Department ELEC, Pleinlaan 2, B1050 Brussels, Belgium.}
and~A.~Moschitta~\IEEEmembership{Member,~IEEE}}

\maketitle
\copyrightnotice
\begin{abstract}
\boldmath
The estimation of the parameters of a dynamic signal, such as a sine wave,
based on quantized data, is customarily performed using the least-square estimator 
(LSE), such as the sine fit. 
However,  the characteristic of the  experiments and of the measurement setup  hardly 
satisfy the requirements ensuring the LSE to be optimal in the minimum mean-square-error sense.
This occurs  if the input signal is characterized by a large signal-to-noise ratio resulting 
in the deterministic component of the quantization error dominating the random error component, {\color{black}and}
when the ADC transition levels are not uniformly distributed over the quantizer input range.

In this paper, it is first shown that the LSE applied to quantized data does not perform as expected when the quantizer is not uniform.
Then, an estimator is introduced that overcomes these limitations. 
It uses the values of the transition levels so that a prior quantizer calibration phase is necessary.
The estimator properties are analyzed and both numerical and experimental results are described to illustrate its performance. It is shown that the described estimator outperforms the LSE and it also 
provides an estimate of the probability distribution function of the noise before quantization.  

\end{abstract}


\newcommand{\fg}[1]{{\frac{1}{\sqrt{2\pi}\sigma} e^{-\frac{{#1}^2}{2\sigma^2}}}} 

%
\IEEEpeerreviewmaketitle


\section{Introduction}
When measuring  the parameters of a noisy signal using quantized 
data, often the least-square estimator (LSE) is used. 
Accordingly, the parameters of the input signal are estimated by 
choosing those values
minimizing the squared error between the input and quantizer output signals.
This is the case, for instance, 
when an analog-to-digital converter (ADC) or a 
waveform digitizer {\color{black}is} tested using the procedures described in 
\cite{Std1241, Std1057}.
The LSE  is known to be optimal under Gaussian experimental conditions. However, this
is rarely the case when data are quantized by a {\color{black}memoryless} ADC, 
{\color{black} unless the input signal is characterized by a low signal-to-noise ratio (SNR).}

Moreover, even if the transition levels in the used quantizer are uniformly distributed over the ADC input range, 
the LSE  is known to be biased \cite{CarboneSchoukens,Alegria,Handel1} and sensitive 
to influence factors such as harmonic distortion and noise \cite{Deyst}. Modifications of the original algorithm that overcome some of these limitations were proposed in \cite{KollarBlair}. 
In practice{\color{black},} however, transition levels are not uniformly distributed in an ADC and 
the LSE or its modified versions produce suboptimal results.

If the values of the ADC transition levels are known, 
the input signal parameters {\color{black}can be estimated}
with better accuracy than the LSE. 
This knowledge is used for instance by maximum-likelihood estimators applied to quantized data \cite{Kollar2}, whose main limitation is the `curse of dimensionality' \cite{ChandrasekaranJain}.
{\color{black} Moreover,} they rely on numerical calculations that may result in suboptimal estimates due to local minima in the cost function.
Alternative estimators based
on sine wave test signals were recently published in \cite{Handel2014} 
to measure specifically the {\color{black}SNR} in an ADC,
showing the ongoing interest of the instrumentation and measurement community to this topic.  

Several results are published about estimators using {\em categorical} data as those 
output by ADCs. A general discussion within a statistical framework can be found in \cite{Agresti}, 
where the usage of {\em link} functions applied to ordinal data is described.     
References \cite{WangYinZhangZhao,104} contain an extensive description of estimators applied to quantized data 
and of their asymptotic properties, 
mainly in the context of system identification and control.
In \cite{Giaquinto2}, a maximum-likelihood estimator is proposed for static testing
of ADCs using {\em link} functions.       

By extending the results presented in \cite{CarboneSchoukensMoschitta}\
this paper introduces an estimator of the 
parameters of a signal quantized by a noisy ADC denominated Quantile-Based-Estimator (QBE).
The main {\color{black}idea} is that an ADC can first be calibrated
by measuring its transition levels and then used to measure the input signal and noise parameters.

When compared to the LSE
that is customarily used for the estimation of sine wave parameters based on quantized data,
it offers several advantages: a reduced bias when the signal-to-noise ratio is large and a reduced mean-square-error (MSE) when the ADC is not uniform. 
Moreover, it also provides an estimate of the input noise standard deviation and of its cumulative (CDF) and probability density functions (PDF). Estimates are obtained by matrix operations so that the curse of dimensionality issue is avoided.
The QBE operates both when the input signal frequency is known or unknown and with or without synchronization between signal and sampling frequencies. Thus, it advances results presented in 
\cite{CarboneSchoukensMoschitta}, where a similar estimator was applied only in the 
case of known synchronized signal and sampling frequencies.
{\color{black}While results can be used in the context of ADC testing the estimator is applicable 
whenever parametric signal identification based on quantized data is needed. }
At first{\color{black},} a motivating example is illustrated. Then, the estimator is described and its properties analyzed through both simulation and experimental results.


\begin{figure}
\begin{center}
\includegraphics[scale=0.4]{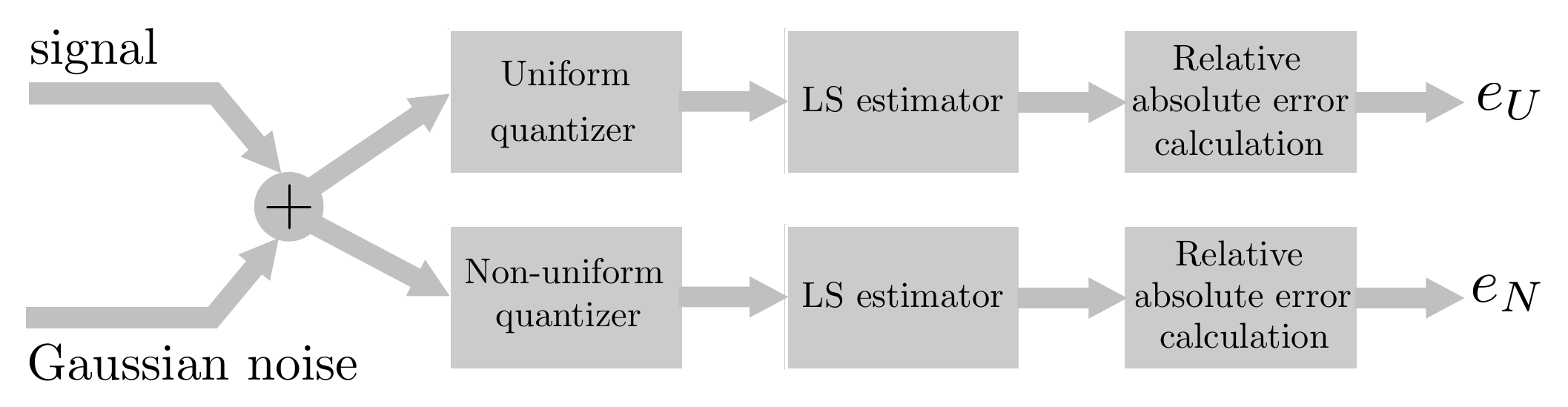}
\caption{The signal chain assumed for showing the effects of INL in the estimation of
the amplitude of a cosinusoidal sequence when a noisy quantization is performed.\label{figmotivating}}
\end{center}
\end{figure}  

\begin{figure}
\begin{center}
\includegraphics[scale=0.4]{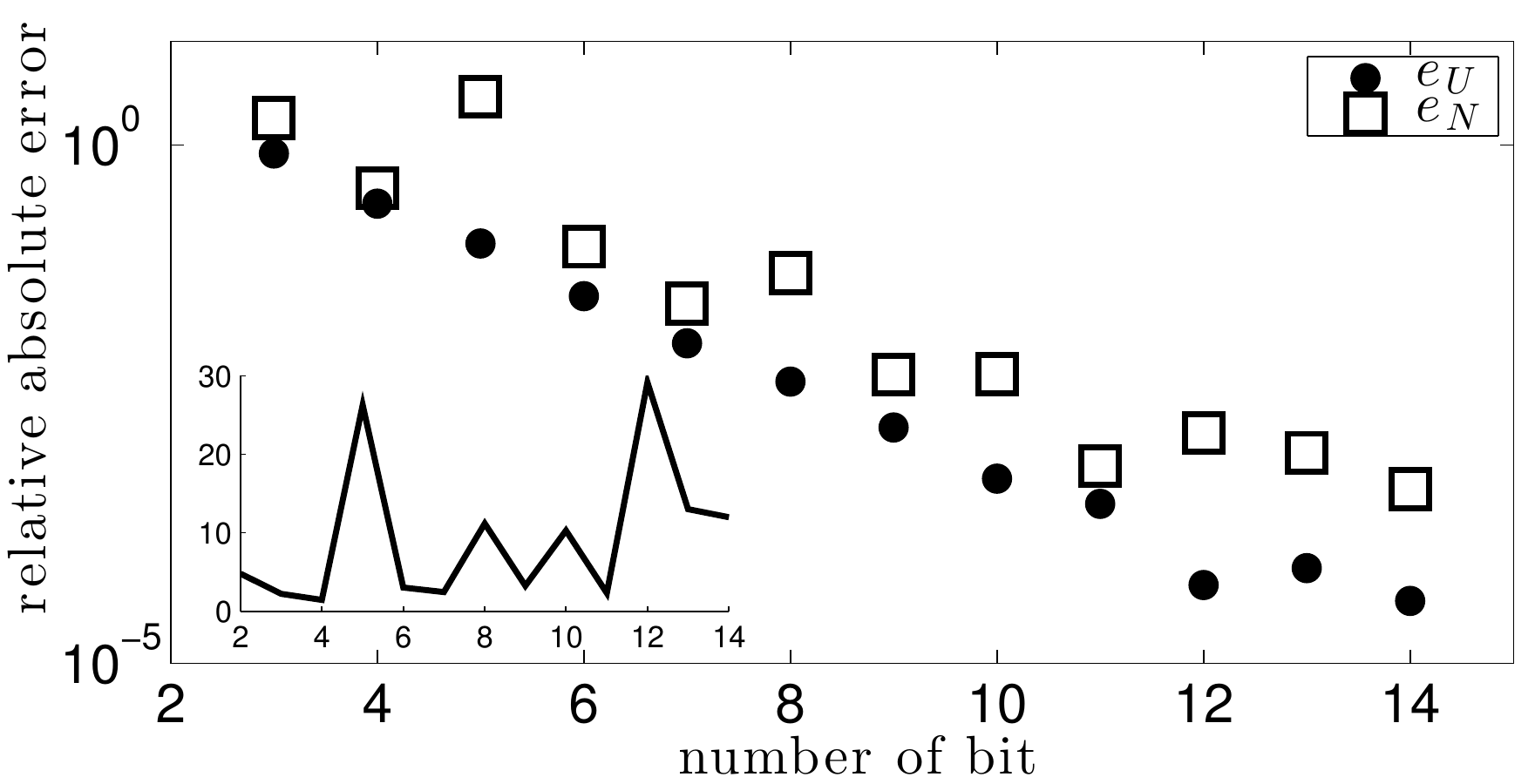}
\caption{{\color{black}Magnitude of the relative} error in the estimation of the amplitude of a cosinusoidal sequence, based on a LSE following the signal chain described in Fig.~\ref{figmotivating}, when the quantizer is both uniform (filled circles) and non-uniform (squares). {\color{black}The inset shows} the ratio between the {\color{black}magnitudes} of  the mean estimation errors. \label{motres}}
\end{center}
\end{figure}

\section{Research Motivation}
Processing of  samples converted using non-uniform quantizers 
requires usage of suitable procedures to extract the maximum possible information from quantized data,
as illustrated in the following subsections.

\subsection{An example}
To show the effect of a non-uniform distribution of transition levels in an ADC when estimating  the amplitude $A$ of a cosine signal by means of the LSE, consider the sequence
\begin{equation}
	x_n = A \cos\left( 2\pi \lambda_n\right) \quad \lambda_n = 10 \frac{n}{N}, \qquad n=0, \ldots, N-1 
	\label{ex}
\end{equation} 
where $0< A \leq 1$ and $N$ is the number of collected samples.
Further{\color{black},} assume that the sequence is affected by zero--mean 
additive Gaussian noise with standard deviation $\sigma=0.3 \Delta$, where $\Delta=\nicefrac{2}{2^b}$ and 
$b$ is the number of quantizer bits. 
By processing the noisy data sequence after the application of a rounding $b$-bit quantizer,   
$A$ is estimated through an LSE when the quantizer is both uniform and non-uniform. 
In this latter case, transition levels are assumed displaced by their nominal position, each by a random variable
uniformly distributed in $\left[ -0.45\Delta, 0.45\Delta \right]$ to introduce integral nonlinearity (INL) while maintaining monotonicity of the input/output characteristic. Both signal chains are shown in Fig.~\ref{figmotivating}.  

The relative absolute errors $e_U$ and $e_N$ in the uniform and non-uniform cases, respectively, are considered as estimation quality criteria.
Results obtained by simulating the signal chains shown in Fig.~\ref{figmotivating} with $A=2^{b-2}\Delta+\nicefrac{\Delta}{2}$ and by collecting $100$ records 
of $N= 10^4$ samples each, are shown in  Fig.~\ref{motres} using a semilogarithmic scale. 
{\color{black}The inset shows} the ratio between the {\color{black}magnitudes of the} estimation errors {\color{black} in the non-uniform and uniform case, respectively}.
Observe that the INL always results in worse performance and in a ratio between 
the {\color{black}magnitudes of the} mean errors as large as $29$. This is not surprising, as INL destroys the otherwise periodic behavior of the quantization error input-output characteristic and because its effects are only marginally attenuated by 
the addition of noise.    

\subsection{An improved approach}
The loss in estimation performance highlighted in Fig.~\ref{motres} arises because the LSE processes the quantizer output codes corresponding to a specific quantization bin. However, while the bin width is constant
in a uniform quantizer, it changes from bin to bin in a non-uniform quantizer, resulting in the so-called differential nonlinearity (DNL). 
Processing data in the code domain does not acknowledge this difference, 
as codes already embed the associated errors. Consequently{\color{black},} information loss is expected.
{\color{black}As an alternative,
data can be processed in the amplitude domain so to avoid 
usage of quantizer codes.}
This approach is feasible if the values of the transition levels in the used quantizer are known or, equivalently, 
are measured before ADC usage. 
Observe that knowledge of transition level{\color{black}s}  allows usage of maximum-likelihood estimators as in 
\cite{Kollar2}. However, this may result in a high computational load and 
in the need to neglect suboptimal solutions, when numerically maximizing the likelihood function. 
Instead, the described technique is based on matrix computation and does not require iterated
numerical evaluations, when the 
input signal frequency is known. By using results published in \cite{CarbonePDF} it will be shown that this procedure also provides an estimate of the input noise standard deviation and of its CDF.
  
\section{Quantile-based estimation \label{section:QBE}}


The main idea of quantile-based estimation was described  
in \cite{CarboneSchoukensMoschitta} and it is here exemplified to ease interpretation of mathematical derivations. 
\subsection{{\color{black}The estimator working principle}}
Assume that the measurement problem 
consists in the estimation of 
an unknown constant value $\mu$, affected by zero-mean additive 
Gaussian noise with known standard deviation $\sigma$.
Consider also the case 
in which the noisy input is 
quantized repeatedly by a comparator with known threshold $T_0$ {\color{black}that}
outputs $0$ and $1$, {\color{black} if the input is below or above $T_0$, respectively.} 
By repeating the experiment several times, the probability of 
collecting samples with amplitude lower than the threshold can be estimated by the percentage count $\hat{p}_0$ of 
the number of  times $0$ is observed.   
This probability can be written as:
 \begin{equation}
 	p_0 = \Phi\left( \frac{T_0-\mu}{\sigma}\right)
	\label{p0}
 \end{equation}
 where $\Phi(\cdot)$ is the CDF of a standard Gaussian random variable. 
Thus, by inverting (\ref{p0})  and substituting $\hat{p}_0$ for $p_0$, 
we obtain the estimate $\hat{\mu}$ of
$\mu$ as
\begin{equation}
	\hat{\mu} = T_0-\sigma\Phi^{-1}(\hat{p}_0),
	\label{firstinversion}
\end{equation}
where all the rightmost terms are known.
Similar arguments can be invoked to solve estimation problems when the quantizer is multi-bit,
when the input sequence is time-varying, it is synchronously or asynchronously sampled and when removing the hypotheses about knowledge of the noise standard deviation. 

\begin{figure}
\begin{center}
\includegraphics[scale=0.1]{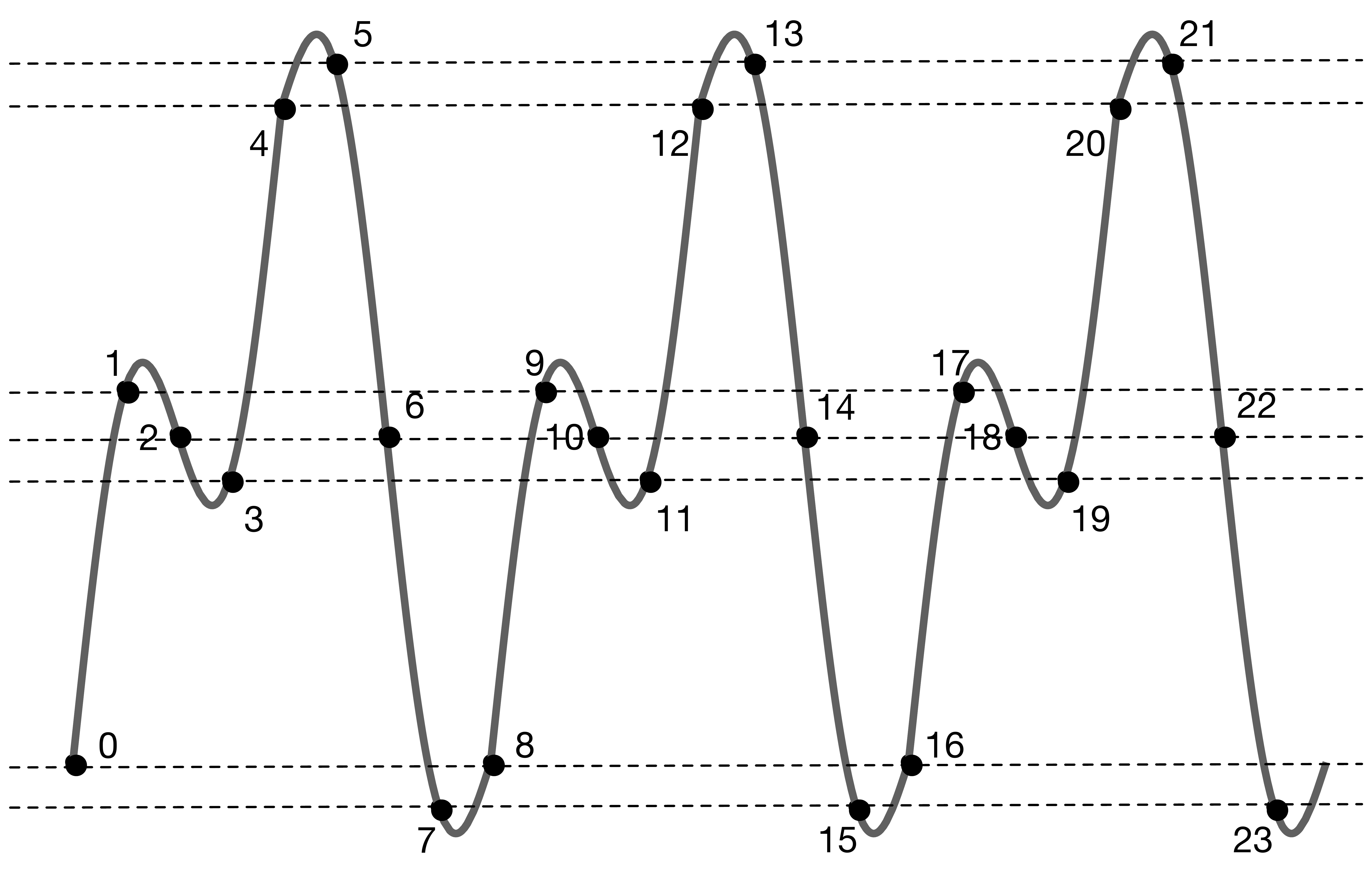}
\caption{A periodic signal ${\color{black} x_P(t)}$ sampled synchronously at $8$ samples per period that is also the period
of the resulting sampled sequence ${\color{black}x[n]}$. 
Dashed lines connect corresponding samples. 
When time indices belong to
any such subset of corresponding samples the probability ${\color{black}p_k[m]}$ can be 
estimated for every possible value of the transition level as shown in Fig.~\ref{probmodel}. 
{\color{black} Here, for every $k$ and every threshold $T_k$, $7$ such probabilities can, in principle, be estimated}.
\label{newmodel}}
\end{center}
\end{figure}

{\color{black} To illustrate the estimation of the 
amplitude of a time-varying signal based on  
noisy quantized samples}, consider the periodic signal shown in Fig.~\ref{newmodel}.
Synchronous sampling of this signal provides the periodic sequence graphed in Fig.~\ref{newmodel} using dots.
By suitably selecting samples within this sequence, 
data can be thought as if they were obtained through {\color{black}the} sampling of $7$ constant values,
graphed in Fig.~\ref{newmodel} using dashed lines. 
The amplitude of the original signal affects the relative distance
among these {\color{black} constant} values. 
{\color{black}
Then, a mathematical model, similar to (\ref{p0}), is used to 
relate the unknown signal amplitude to the values taken by the sampled data. 
Once code occurrence probabilities are estimated 
using the quantizer output sequence, 
knowledge of this model and its inversion, result in the estimation of the
amplitudes of the constant sequences shown in Fig.~\ref{newmodel} 
and of the overall signal, as in (\ref{firstinversion}).}
In the following subsections this 
procedure will be explained in more depth. 
{\color{black}  Accordingly, the next section introduces the signal and system 
models used in the estimation procedure.}

\subsection{Signals and Systems}
For $n=0, \ldots, N-1$ we assume:
\begin{align}
\begin{split}
\Theta & =[ \theta_0 \; \theta_1 \; \cdots \; \theta_{M-1}]^T 	\\	
S[n] & =[ s_0[n] \; s_1[n] \; \cdots \; s_{M-1}[n]] ^T	\\	
x[n] & = S[n]^T\Theta \\
x_q[n] & = Q(x[n]+\eta[n]) 
\label{sigmodel}
\end{split}	
\end{align}
where $x[n]$ is a discrete-time sequence obtained by sampling a periodic continuous-time signal, 
$S[n]$ represents a vector of known discrete-time values $s_0[n], \ldots, s_{M-1}[n]$, and $\eta[\cdot]$ is a sequence 
of independent and identically distributed zero-mean Gaussian random variables, having standard deviation $\sigma$. In (\ref{sigmodel}), $\Theta$ represents the vector of unknown parameters, $Q(\cdot)$ represents the quantization operation and $e[\cdot]$ the associated quantization error sequence. Quantization results {\color{black} in $x_q[n]$ taking} one of the
 $K$ possible ordered quantization codes  {\color{black} $Q_{k-1}$, $k \in {\cal K}=\{1, \ldots, K\},$} if the quantizer input belongs to the interval  {\color{black} $[T_{k-1}, T_{k})$}, where $T_k$ represents the $k$-th quantizer transition level.  
{\color{black}In practice, $x[n]$ is measured in volt, while $x_q[n]$ is coded by the ADC, according to the choice made by the producer (e.g. binary, decimal).} 
 
To exemplify the use of the signal model (\ref{sigmodel}), consider the case described in
subsection A. Accordingly, $M=1$, $s_{\color{black}0}[n] \equiv 1$, 
the output of the comparator $Q(\cdot)$ can be either $0$ or $1$, based on the value 
of the unique threshold level $T_0$ and $N$ still represents the number of observed and processed samples.
Similarly, (\ref{sigmodel}) models (\ref{ex}) by assuming $M=1$, $\theta_0=A$ and $S[n]=s_0[n]=\cos\left( 2\pi \cdot 10 \frac{n}{N}\right)$, obtained by sampling synchronously the signal
$\cos\left( 2\pi f t\right)$, where $f$ is the signal frequency.

{\color{black} Finally, the case shown in Fig.~\ref{newmodel} is modeled by assuming $M=2$ and the sampled sequence $x[n]= \theta_0s_0[n]+\theta_1s_1[n]$, $n=0, \ldots, 23$, 
where $\theta_0$, $\theta_1$ represent the two unknown amplitudes to be estimated
and 
\begin{align}
\begin{split}
	s_0[n]  & = \arccos\left(\cos\left(2\pi \langle 0.125 \cdot n  \rangle \right)\right) \\
	s_1[n]  & =  \sin\left(4 \pi \langle 0.125 \cdot n \rangle \right),
\end{split}
\end{align}
two known sequences, with $\langle x \rangle$ representing the fractional part of $x$.
}

Observe that, in general,
$x[n]$ is itself a periodic sequence if sampling is done synchronously, 
that is the ratio between the signal frequency and sampling rate is a rational number,  and aperiodic otherwise.
 
\subsection{Extension to multi-bit quantizers}

The approach described in subsection A, based on a single threshold $T_0$, 
can be extended to comprehend {\color{black} both} the multi-threshold case 
that applies when using multi-bit quantizers {\color{black} and the case when the input signal is time-varying. The first example in subsection \ref{section:QBE}-A was related to a constant input signal resulting in a constant probability to be estimated, as shown by (\ref{p0}).
When the input signal is time-dependent
the probabilities to be estimated  
are no longer constant, but time-dependent as well. This is indicated in the following by using the symbol $p_k[n] = P(x_q[n] \leq Q_{k-1})$.
}
{\color{black} Then}, for $k \in {\cal K}$ and $n=0, \ldots, N-1$ {\color{black} one can write}
\begin{align}
\begin{split}
P(x_q[n] \leq Q_{k-1}) & = P(x[n]+\eta[n] \leq T_k) \\
 & = P(\eta[n] \leq T_k-x[n]) \\
& = \Phi\left( \frac{T_k-x[n]}{\sigma}\right) = 
\Phi\left( \frac{T_k- S[n]^T\Theta}{\sigma}\right)
\label{modeleq}
\end{split}	
\end{align}

If an estimate $\hat{p}_k[n]$ of $p_k[n]$ 
is available,
such that $0 < \hat{p}_k[n] < 1$, 
 (\ref{modeleq})
can be inverted according to whether $\sigma$ is known or unknown. 
If $\sigma$ is known, from (\ref{modeleq}) we have:
\begin{align}
\begin{split}
\Phi^{-1}\left( \hat{p}_k[n] \right) = \frac{T_k-S[n]^T\Theta}{\sigma}.
\label{eqmodel}
\end{split}	
\end{align}
that is, for $k \in {\cal K}$ and $n=0, \ldots, N-1$
\begin{align}
\begin{split}
S[n]^T\Theta = T_k-\sigma\Phi^{-1}\left( \hat{p}_k[n] \right), 
\label{eqfund}
\end{split}	
\end{align}
If $\sigma$ is unknown, it must be estimated 
{\color{black} using the} quantized data, so that from (\ref{eqmodel}) we have:
\begin{align}
\begin{split}
S[n]^T\frac{\Theta}{\sigma} - \frac{T_k}{\sigma}= -\Phi^{-1}\left( \hat{p}_k[n] \right), 
\end{split}	
\end{align}
that, for $k  \in {\cal K}$ and $n=0, \ldots, N-1$ results in
\begin{align}
\begin{split}
\left[ S[n]^T \; T_k\right]\Theta_U = -\Phi^{-1}\left( \hat{p}_k[n] \right), 
\label{eqfundu}
\end{split}	
\end{align}
in which, by setting $\theta_M=\sigma$, $\Theta_U$ is defined as 
\begin{equation}
\Theta_U=\left[ \frac{\theta_0}{\theta_M} \; \frac{\theta_1}{\theta_M} \; \cdots \; \frac{\theta_{M-1}}{\theta_M}\; \frac{-1}{\theta_M}\right]^T.\\
\end{equation}
Notice  that 
if an estimate {\color{black}$\hat{\Theta}_U$} of $\Theta_U$ is available, the {\color{black}scalar} parameters $\theta_0, \ldots, \theta_M$ can 
be recovered by means of simple transformations of the elements in {\color{black}$\hat{\Theta}_U$}. 
{\color{black}The noise standard deviation $\sigma=\theta_M$ 
can be estimated by inverting the rightmost element in {\color{black}$\hat{\Theta}_U$} and by changing its sign.
Estimates of $\theta_0, \ldots, \theta_{M-1}$ are obtained
by multiplying the $M$ leftmost elements in {\color{black}$\hat{\Theta}_U$} by the obtained value of $\theta_M$.}

\subsection{The derivation of the proposed estimator}
Expressions (\ref{eqfund}) and (\ref{eqfundu}) 
show a linear relationship between the vector of unknown parameters 
and the estimated probability ${\color{black}{\hat p}_k[n]}$, suitably transformed by the so--called
{\em link} function $\Phi^{-1}(\cdot)$ \cite{Agresti, WangYinZhangZhao}.
Observe that, in principle, $K\cdot N$ such relationships are available. In practice{\color{black},} however, the number is
much lower for two reasons:
\begin{enumerate}
\item the application of the link function requires the inversion of $\Phi({\color{black} p_k[n]})$ that is unfeasible if the estimated probability is equal to $0$ or $1$. When this occurs, data are discarded;
\item 
{\color{black} for a given $T_k$, the estimation of a probability requires a percentage count of the codes that result in the noisy input signal being below or equal to $T_k$. This would require the input signal $x[n]$ to remain constant over $n$, while several such codes are collected.
Since this is unfeasible when the signal is time-varying unless the input sequence is largely oversampled, there is the need to identify subsets of time indices $n$ approximately providing the same value of $x[n]$. 
Accordingly, the $N$ time indices are partitioned into subsets associated with 
values of $x[n]$ having close magnitude. Consequently, the estimation of a single probability requires usage of several input samples resulting in a number of estimates that for a given $k \in {\cal K}$ is lower than $N$. The partitioning mechanism will be presented in the next subsection.}   
\end{enumerate}

\begin{figure}
\begin{center}
\includegraphics[scale=0.35]{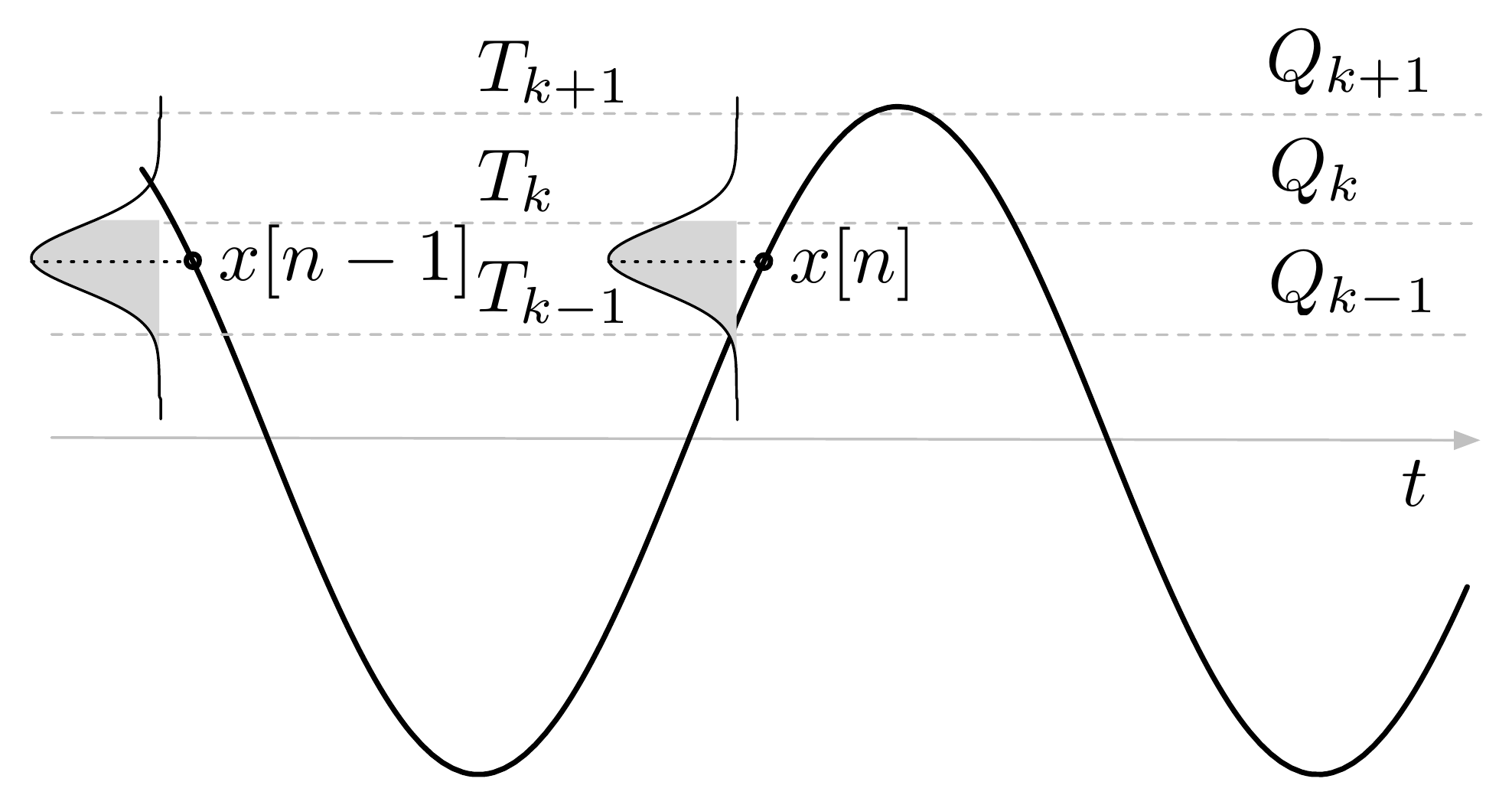}
\caption{{\color{black}Estimation} of the probabilities ${\color{black} p_k[m]}$: for a given subset of indices 
$\{n-1, n\}$, $x[n]$ takes the same value $x_{n-1}=x_n$ and the shaded probability can be estimated by 
counting the number of times 
the code is less than or equal to $Q_k$ and dividing by $2$. Observe that in this very simple case the only possibility for ${\color{black} p_k[m]}$ not to be equal to $0$ or $1$, is when the number of counts is equal to $1$. The other two cases would be discarded by the algorithm as the {\em link} function $\Phi^{-1}({\color{black} p_k[m]})$ would not be invertible.
\label{probmodel}}
\end{center}
\end{figure}  

\subsection{Estimation of probabilities ${\color{black} p_k[n]}$}
The determination and inversion of the measurement model as in 
(\ref{firstinversion}) and in (\ref{eqfundu})
require knowledge of probabilities ${\color{black} p_k[n]}$, as defined in (\ref{newmodel}).  
{\color{black} In general, if $x[n]$ is obtained by sampling a periodic signal $x_P(t)$ with period $T$, 
one obtains
      $ x[n] = x_P(nT_s) = x_P\left( \left \langle \frac{nT_S}{T} \right \rangle T\right)$.
{\em Synchronous sampling} applies
if $\nicefrac{T_S}{T}$ is a rational number 
resulting in a periodic sequence $x[n]$.
When $\nicefrac{T_S}{T}$ is irrational, sampling is {\em asynchronous} and
$x[n]$ is no longer a periodic sequence.
The two cases are treated separately in the following.

\subsubsection{{\em Synchronous sampling} (rational $\nicefrac{T_s}{T}$)\label{subsync}}
If $\frac{T_s}{T} = \frac{L}{N}$
where $L$ is an integer number, 
       $x[n] =  
       x_P\left( \frac{nL\bmod N}{N}  T\right),$
results, 
where ${\bmod}$ is the remainder operator. }
It is of interest to analyze the image of the map $I_L=(nL\bmod N)$ when $n=0, \ldots, N-1$.
By the theorem 2.5 in \cite{Shoup}, this image consists of the 
$\nicefrac{N}{d}$ integers 
\begin{equation}
	n\cdot d, \qquad n=0, \ldots, \frac{N}{d}-1
\label{theo}
\end{equation}
where $d$ is the greatest common divider of $L$ and $N$.
By this argument, $x[n]$ can only take the values
\begin{align}
\begin{split}	
       x[n]=x_P\left(  \frac{nd}{N}  T\right),  \qquad n=0, \ldots, \frac{N}{d}-1
\label{perseq3}
\end{split}	
\end{align}
which may not be all unique, and $x[n]$ becomes a periodic sequence with period given by $\frac{N}{d}$. As an example, if $N=10$ and $L=1$, $d=1$ results, the map $I_L$ provides the unique values
$n=0, \ldots, 9$, and the samples represent a single period of ${\color{black}{\color{black} x_P(t)}}$. 
Conversely, if $N=10$ and $L=2$, $d=2$ and the image of the map $I$ 
contains the values $2n$, $n=0, \ldots, 4$. These values are provided twice when $n=0, \ldots, 9$ so that
the sequence ${\color{black} x[n]}$ has period $\nicefrac{10}{2}=5$.
Each one of the $\frac{N}{d}$ different values in the image of the map is repeated $d$ times.
Each time, this value results in the same value of the signal $x[n]$ provided to the ADC.

{\color{black}T}he synchronous case is exemplified in Fig.~\ref{newmodel}, where $3$ periods of a synchronously sampled periodic signal ${\color{black} {\color{black} x_P(t)}}$ are shown,  when $N=24$, $L=3$, so that $d=3$ results. Samples corresponding to the same input value are connected through dashed lines. 
Thus a partition ${\cal P}=\{ \{0, 8,16\}, \{1, 9,17\}, \{2,10,18\}, \{3,11,19\}, \{4,12,20\}, 
\{5,$ $13, 21\},\{6,14,22\},
\{7,15,23\}\}$ of the indices $0, \ldots, 23$ is obtained.
Then,
for any given $k$ identifying the selected threshold and for every 
${\color{black}m= 0, \ldots, |{\cal P}|-1}$, where 
$|{\cal P }|$ represents the cardinality of the partition,
the samples belonging to each subset {\color{black}$P_m$ in} the partition ${\cal P}$ describe the same event 
and a counter can be updated with $1$ or $0$ according to whether or not the quantizer output is less than or equal to $Q_k$, as defined in (\ref{modeleq}).  
Thus, {\color{black} for a given and known threshold $T_k$, $p_k[m]$, $m = 0, \ldots, |{\cal P}|-1$} can be estimated by the percentage count accumulated over those indices {\color{black} in $P_m$} providing the same 
quantizer input $x[n]$. Even though the exact value of $x[n]$ is not known by the user, as it depends on the unknown
parameter values, if the same argument $t$ of $x_P(t)$ repeats over time, so does its sampled version $x[n]$. 
Thus, by matching the time instants corresponding to the same value of ${\color{black} x_P(t)}$, an estimate ${\color{black} p_k[m]}$ can be obtained for every $k$. This is exemplified in Fig.~\ref{probmodel}, where a sinusoidal signal ${\color{black} x_P(t)}$ is assumed
and the probability to be estimated is shaded. In this figure, the values of $3$ transition levels are shown using dashed lines and the corresponding quantizer codes are indicated as $Q_{k-1}$, $Q_k$ and $Q_{k+1}$.
Results on the application of synchronous sampling to the estimation of the parameters of a sinusoidal function are published in \cite{CarboneSchoukensMoschitta}.

\begin{figure}
\begin{center}
\includegraphics[scale=0.12]{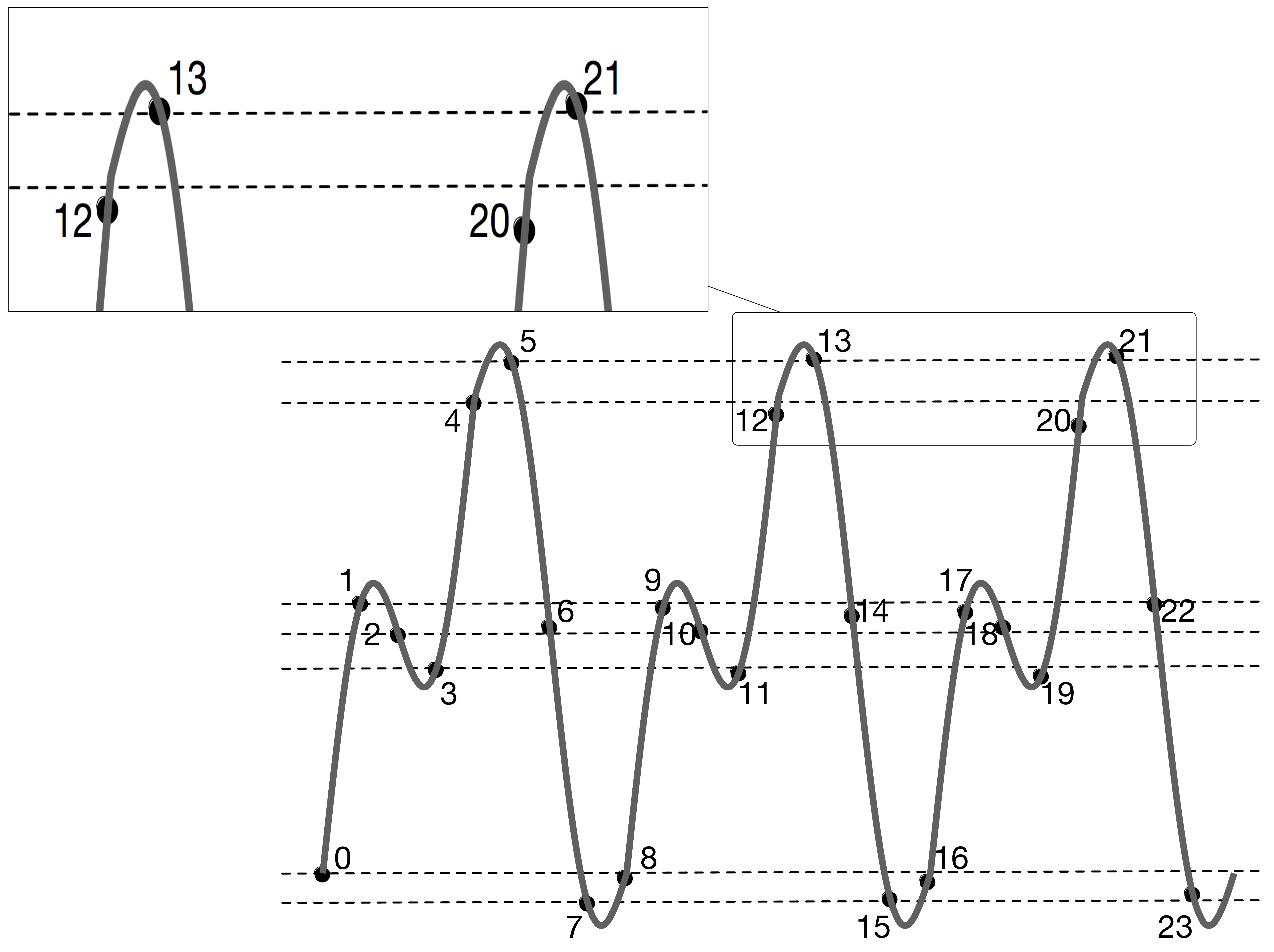}
\caption{A periodic signal ${\color{black} x_P(t)}$ sampled asynchronously at about $8$ samples per period 
resulting in an aperiodic sampled sequence $x[n]$. 
Dashed lines connect corresponding samples. 
Samples can be grouped according to their closeness so that an estimate of the probability for the signal
to be above or below a given threshold can then be estimated (see text). 
{\color{black} The enlarged detail shows the effect of asynchronous sampling to the relative position of samples, in regions of the signal exhibiting different time derivatives.}
\label{newmodela}}
\end{center}
\end{figure}  

\subsubsection{{\em Asynchronous sampling} (irrational $\nicefrac{T_s}{T}$)}
In practice, 
synchronization requires {\color{black}the} careful setup of the experiments
since disturbing mechanisms may occur.  
For instance, basic bench top equipment used to generate signals 
is affected by frequency drifts over time that can not easily be compensated for.
Similarly, the used ADC may sample inputs 
with a sampling period that may vary over time and that may not be controlled directly by the user or 
disciplined by external stable sources such as cesium or rubidium frequency standards.
As a consequence{\color{black},} the ratio between sampling and signal periods is more properly assumed as 
\[
	\frac{T_s}{T} = \lambda,
\]
where $\lambda$ is an irrational number and
\begin{equation}
	x[n] = x_P\left (\left\langle n \lambda \right \rangle T \right), \qquad n=0, \ldots, N-1
\label{irra}
\end{equation}
becomes an aperiodic sequence.  
This case is graphed in Fig.~\ref{newmodela}, where $3$ periods of an asynchronously sampled periodic signal ${\color{black} x_P(t)}$ are shown,  when $N=24$ and $\lambda=0.1245\cdots$. Reference dashed lines show
the progressive deviations from synchronicity when the time index increases.   
Contrary to the rational case, 
since the map ${\color{black} I_{\lambda}} = \langle n \lambda \rangle$ has no periodic orbits, when $n=0, \ldots, N-1$, no two equal values of the argument of ${\color{black} x_P(t)}$ in (\ref{irra}) can be found. 

Remind that probabilities $p_k[{\color{black}m}]$ can be estimated only when the same value of $x[n]$ is input to the ADC.
However, when $\lambda$ is irrational, $x[n]$ provides only approximately equal values for selected
indices $n$. Accordingly, by choosing 
a small value $\epsilon>0$, the image of ${\color{black} I_{\lambda}}$ can be explored 
to find those indices $n$ for which ${\color{black} I_{\lambda}}$ returns values that differ at most by $\epsilon$.
If  the sequences in $S[n]$ have bounded derivatives, 
small deviations in their arguments will result in bounded variations of their amplitudes
and the estimation will occur as if synchronous sampling was adopted.
The main idea is that by selecting arguments for ${\color{black} x_P(t)}$ in (\ref{irra}) 
that are close to each other, ${\color{black} x_P(t)}$ will result
in samples with similar values, at the same time.
Thus, the estimation procedure can be described by the following steps:
\begin{enumerate}
\item the interval $[0,1)$ is partitioned {\color{black}into} adjacent subintervals, each of length $\epsilon>0$;
\item each subinterval is associated with that particular set of indices $n$ for which ${\color{black} I_{\lambda}}$ 
returns values belonging to that interval. The collection of these sets represents a partition 
${\cal P}$
of the whole set of integers $n=0, \ldots, N-1$;
\item for every {\color{black}subset $P_m$ in this partition}, 
a new vector ${\color{black} \overline{S}[m]=[ \overline{s}_0[m] \; \overline{s}_1[m] \; \cdots \; \overline{s}_{M-1}[m]]^T}$
is defined, where each sequence ${\color{black}\overline{s}_i[m]}$ 
is obtained by averaging ${\color{black}s_i[m]}$ over the indices in ${\color{black}P_m}$. 
This vector takes the role of ${\color{black}S[m]}$ in (\ref{eqfund}) and (\ref{eqfundu}). Corresponding estimates of
${\color{black} p_k[m]}$ are obtained by accumulating counts when these indices occur.
\end{enumerate}  

{\color{black}T}he asynchronous case is exemplified in Fig.~\ref{newmodela}. Assuming $N=24$ and $\lambda = 0.1245\cdots$, the sampled sequence can be modeled as:
\begin{align}
\begin{split}
	x[n] = \theta_0\arccos\left(\cos\left(2\pi \langle n \lambda \rangle \right)\right)+
	& \theta_1\sin\left(4 \pi \langle n \lambda \rangle \right),\\
	&  \qquad n=0, \ldots, 23
\end{split}
\end{align}
Using the notation defined in (\ref{sigmodel}), $M=2$ and we can {\em initially} write:
\begin{align}
\begin{split}
\Theta & =[ \theta_0 \; \theta_1 ]^T 	\\	
S[n] & =[ \arccos\left(\cos\left(2\pi \langle n \lambda \rangle \right) \right) \; \sin\left(4 \pi \langle n \lambda \rangle \right) ] ^T	\\	
x[n] & = S[n]^T\Theta \\
\end{split}	
\end{align}
By assuming $ \epsilon=0.1$, the procedure returns
the partition of the set of indices ${\cal P} =  
\{ \{0\}, \{ 1,9,17\}, \{ 2,10,18\}, \{3,11,19\},$ $\{4,12,20\}, \{5,13,21\},\{ 6,14,22\}, \{7,15,23\},\{8,16\} \}$, where {\color{black}$|{\cal P}|=9$ represents a bound on the number of probabilities that can be estimated for every $k$. 
The actual number might be lower because of the additional constraint $0 < \hat{p}_k[m] < 1$}. 
{\color{black}Subsets $P_m \in {\cal P}$} identify samples of ${\color{black} x[n]}$ having approximately the same magnitude. This is shown 
in Fig.~\ref{newmodela} where $\theta_0=\theta_1=1$ is assumed and each sample is identified by the corresponding value of $n$.
{\color{black} Asynchronous sampling results in different displacements among corresponding samples because of the different derivative of the signal in different temporal regions. An enlarged detail in Fig.~\ref{newmodela} shows this phenomenon.}

Observe that estimates of ${\color{black} p_k[m]}$ that differ from $0$ and $1$ require subsets with at least $2$ 
indices, as at least $2$ counts are needed. 
Thus, for every  $P_m \in {\cal P}$  
and for every possible transition level $T_k$, a corresponding probability ${\color{black} p_k[m]}$ can be estimated.
In this example out of the available $24$ samples, only $8$ sets are available for estimating corresponding sets of probabilities ${\color{black} p_k[m]}$,
when $k=1, \ldots, K$. 
Accordingly, for every ${ P}_{\color{black}m} \in {\cal P}$  and for every $k=1, \ldots, K$ 
the corresponding value of the known sequence 
can {\em finally} be written as in the model:
\begin{align}
\begin{split}
\Theta & =[ 1 \; 1 ]^T 	\\	
\overline{S}[m] & =[\overline{s}_0[m] \; \overline{s}_1[m]]^T \\
& \hskip-0.5 cm= \left[ \frac{1}{N_{\color{black} m}}   \sum_{n \in { P}_{\color{black} m}}\arccos\left(\cos\left(2\pi \langle n \lambda \rangle \right) \right) \; \frac{1}{N_{\color{black} m}}\sum_{n \in { P}_{\color{black} m}}\sin\left(4 \pi \langle n \lambda \rangle \right) \right] ^T	\\	
x[m] & = \overline{S}[m]^T\Theta,{\color{black} \qquad m=0, \ldots,  |{\cal P}|-1} \\
\label{sigmodelex}
\end{split}	
\end{align}
{\color{black} where $N_{\color{black}m}$ represents the cardinality of $P_m$} {\color{black}and} 
the bar reminds that the known signals $\overline{s}_0[{\color{black} m}]$
and $\overline{s}_1[{\color{black} m}]$ are obtained after averaging all 
approximately equal amplitude values associated {\color{black} with} indices in ${P}_{\color{black} m}$.

{\color{black} \subsection{Model inversion and parameter estimation}
Define ${\cal S}$ as the set containing {\color{black} only}
couples of indices $(k,m)$, 
{\color{black} allowing estimation of $p_k[m]$, that is implying $0<\hat{p}_k[m]< 1$}. Then, (\ref{eqfund}) can be put in matrix form as follows. 
When $\sigma$ is known, for each couple of indices $(k,m) \in {\cal S}$
\begin{itemize}
\item  a row is added to a matrix
$H$ containing the vector $S[m]$ or $\overline{S}[m]$, in the case of synchronous or asynchronous sampling respectively;
\item a row is added to a column vector $Y$ containing the scalar
$T_k-\sigma\Phi^{-1}(\hat{p}_k[m])$, where $T_k$ and $\sigma$ are known and 
$\hat{p}_k[m]$ is estimated using the data, 
as shown above. 
\end{itemize}
Once all indices $(k,m)$ in ${\cal S}$ are considered, 
the linear system 
\begin{equation}
	H\Theta = Y,
\label{linear1}
\end{equation}
results. Observe that, by construction, the number of rows in $H$ and $Y$ is a random variable as 
 ${\cal S}$ contains a random number of entries.
Then, if the number of rows in $H$ is not lower than the number of unknown parameters,
an estimate of $\Theta$ can be obtained by applying a least-square estimator as follows:
\begin{equation}
	\hat{\Theta} = (H^TH)^{-1}H^TY.
\label{teta}
\end{equation}
Similarly when $\sigma$ is unknown, for each couple of indices $(k,m)$ in ${\cal S}$
\begin{itemize}
\item a matrix $H_U$ can be constructed by adding entries containing the vector 
$S[m]-T_k$ 
or $\overline{S}[m]-T_k$, in the case of synchronous or asynchronous sampling respectively;
\item a column vector
$Y_U$ is created, whose entries are the corresponding values $-\Phi^{-1}(\hat{p}_k[m])$. 
\end{itemize}
The linear system
\begin{equation}
	H_U\Theta_U = Y_U,
\label{linear2}
\end{equation}
results, where  $H_U$ and $Y_U$ have again a random number of rows. 
Finally, $\Theta_U$ can be recovered by 
 a least-square approach as follows:
\begin{equation}
	\hat{\Theta}_U = (H_U^TH_U)^{-1}H_U^TY_U.
\label{tetaU}
\end{equation}
Observe that several techniques can be applied to find an estimator of $\Theta$  and 
$\Theta_U$, starting from (\ref{linear1}) and (\ref{linear2}), respectively.  
As an example, by estimating the covariance matrix associated {\color{black}with} available data, a
weighted least-square 
estimator can be applied, as done in \cite{CarboneSchoukensMoschitta}. 
In this paper, the simplest possible approach based on the application of the least-square solution is taken.}
Finally, observe that the procedure described in this subsection can be applied 
irrespective of the rationality or irrationality of the ratio $\nicefrac{T_s}{T}$. In the former case and for sufficiently small values of $\epsilon$, it will provide  the same set of indices that the user would select by following the indications in section~\ref{subsync}.

\section{A new {\em Sine Fit} procedure}
Fitting the parameters of a sine wave to a sequence of quantized data is a common problem when testing systems, e.g. ADCs or other nonlinear and linear systems.
The LSE is the technique adopted in this case. However, this estimator is known:
\begin{itemize}
\item to be a biased, not necessarily asymptotically unbiased, estimator \cite{CarboneSchoukens};
\item to perform poorly when the resolution of the quantizer is low, e.g. 4-5 bits, and the added noise has a small standard deviation so that the ADC can hardly be considered as a linear system adding white Gaussian noise.    
\end{itemize} 
It will be shown in this section how to use the QBE to obtain an alternative estimator that outperforms the 
LSE with respect to both bias and MSE and both when the sine wave frequency is known and unknown. 
The general case of an irrational value of $\lambda = \frac{T_s}{T}$ is treated in the following since it also includes
the case when $\frac{T_s}{T}$ is rational. The further general assumption of $\sigma$ unknown is considered. 
Two further cases apply: when $\lambda$ is known or unknown
to the user, so that an equivalent formulation of the $3$- or $4$-parameter
{\em sine fit} is obtained, respectively \cite{Std1241}.
 
\subsection{Known Frequency Ratio $\lambda$}
This is the case when the input signal can be modeled as:
\begin{equation}
	x[n] = \theta_0\sin \left(2 \pi \left \langle n \frac{T_s}{T} \right \rangle \right) +
	\theta_1\cos \left(2 \pi \left \langle n \frac{T_s}{T} \right \rangle \right) +\theta_2
\label{modelsine}
\end{equation}
By following the procedure described in section \ref{section:QBE}
in the case of unknown $\sigma$, a small value is chosen for $\epsilon$ that results in the corresponding 
partition ${\cal P}$ of the set of indices $n=0, \ldots, N-1$. 
For every couple
of $(k, {\color{black}m})$, $k=0, \ldots, K-1$, ${\color{black}m}=0, \ldots, |{\cal P }|-1$, a probability $p_k[{\color{black}m}]$
is estimated.
If this estimate $\hat{p}_k[{\color{black}m}]$ differs from $0$ and $1$, the following 
$1 \times 4$ row vector is added to the observation matrix $H_U$
\begin{align}
\begin{split}
\overline{S}[{\color{black}m}] & = \\
& \hskip-0.5 cm= \left[ \frac{1}{N_{\color{black}m}}   
\sum_{n \in P_{\color{black}m}}\sin\left(2\pi \langle n \lambda \rangle \right)  \; \frac{1}{N_{m}}\sum_{n \in P_m}\cos\left(2 \pi \langle n \lambda \rangle \right) \; 1 \; T_k \right] 	
\label{newsinmod}
\end{split}	
\end{align}
and the scalar $-\Phi\left( \hat{p}_k[{\color{black}m}]\right)$ is added to the column vector $Y_U$.
In (\ref{newsinmod}), $N_{\color{black}m}$ represents the cardinality of $P_{\color{black}m}$.
Once all couples in $(k,{\color{black}m})$ are considered, an estimate of $\Theta_U$ is found through
(\ref{tetaU}), from which estimates of $\theta_i$, $i=0, \ldots, 3$ can straightforwardly be derived.

\subsection{Unknown Frequency Ratio $\lambda$ \label{ukn}}
Often{\color{black},} the user {\color{black}is unaware of} the exact value of the ratio between signal {\color{black}frequency} and sampling rate. When this {\color{black}occurs}, an iterative approach {\color{black}applies} \cite{Std1241}:
\begin{itemize}
\item $\lambda$ is {\color{black}initially guessed}, e.g. {\color{black}using the procedure} described in $\cite{IFFT,DalletBelegaPetri}$;
\item using this value of $\lambda$, $\Theta_U$ is estimated {\color{black}following} the procedure described in section \ref{section:QBE}
and the MSE is evaluated;
\item the frequency estimate is updated, e.g. by following the golden section search 
algorithm, with the MSE as the goodness-of-fit criterion \cite{golden};
\item the {\color{black}magnitude of the} deviation in the frequency values from one update to the following {\color{black}is chosen as the stopping rule}: if {\color{black} it} is below a user given value $\gamma$, the procedure is stopped. 
\end{itemize} 
As it happens when the LSE is applied iteratively, 
this procedure converges if the initial frequency guess is 
within a given frequency capture range.  The initial guess provided by 
the discrete-Fourier-transform of quantized data, as suggested in \cite{IFFT,DalletBelegaPetri}, proved to be sufficiently accurate in the cases illustrated in the following sections.

\section{Practical Implementation Issues}
The practical implementation of the QBE algorithm shows that 
setting parameters and interpreting results require some caution. In fact:
\begin{itemize}
\item for a given $N$ and when $\lambda$ is irrational, 
if $\epsilon$ decreases the number of subsets in the partition ${\cal P}$ increases, leading to a large number of different estimates of ${\color{black} p_k[m]}$.
However, at the same time the average number of indices in each 
subset of the partition decreases, resulting in a less accurate estimation of each probability ${\color{black} p_k[m]}$.
Thus, the choice of $\epsilon$ is a result of a compromise: 
either few accurate or many rough estimates are processed by the algorithm. Repeated simulations showed that  
$\epsilon$ approximately results in similar MSEs 
for a wide range of values, since the two effects tend to compensate each other. 

An approximated reasoning can explain this behavior. 
Consider the estimator asymptotic accuracy for small values of $\epsilon$. 
Both the number of counts used to estimate $p_k[m]$ and $\mbox{var}(\hat{p}_k[m])$ are 
$\mathcal{O}\left(\frac{1}{\epsilon}\right)$,
 while the variance in estimating $\Theta$ 
 is $\mathcal{O}\left( \frac{1}{\text{number of counts}}\right) \times
 \mbox{var}(\hat{p}_k[m]) =
 \mathcal{O}(\epsilon)\mathcal{O}\left(\frac{1}{\epsilon}\right) = \mathcal{O}(\epsilon^0)$, that is
 independent of $\epsilon$; 
\item the estimation of $\Phi^{-1}({p}_k[m])$ implies
the application of a nonlinear function to the random variable $\hat{p}_k[m]$ obtained through a percentage count.
While $\hat{p}_k[m]$ based on a percentage of the total number of samples satisfying a given rule, is an unbiased estimator of the underlying unknown probability \cite{Devore}, the application of the nonlinear function results in a biased but asymptotically unbiased estimator. 
Three approaches are possible: 
{\color{black}
\begin{itemize}
\item a lower bound is set to discard estimates based on small size samples; 
\item the bias can be estimated and partially corrected for, e.g., by expanding the nonlinear function using a Taylor series
about the expected value ${\color{black} p_k[m]}$ of ${\color{black} \hat{p}_k[m]}$; 
\item the bias magnitude can be bounded. 
\end{itemize}}
In this latter case, the bias can be expected 
to be more severe when ${\color{black} p_k[m]}$ is close to $0$ and $1$, that is where $\Phi^{-1}(\cdot)$ 
has two vertical asymptotes and exhibits a strong nonlinear behavior.
Thus, to reduce the bias in estimating $\Phi^{-1}({p}_k[m])$,  {\em guard} intervals can be set, so that data are processed by the algorithm only if, e.g. $0.05 < {\color{black} \hat{p}_k[m]} < 0.95$.  
\end{itemize}

\section{Validating the Assumption on the Noise PDF} 
The QBE is based on the assumption that the noise CDF is known, as the inverse of this function represents the
{\em link} function needed to apply the main estimator equation (\ref{eqmodel}). This assumption can be tested by 
estimating the input noise CDF and PDF by following the procedure described in \cite{CarbonePDF}. Accordingly, if the transition levels in the ADC, 
are known, as well as the input sequence $x[n]$, a pointwise estimate of the noise CDF 
for any available estimate $\hat{p}_k[m]$, is provided by:
{\color{black}
\begin{equation}
	\hat{F}_{\eta}(T_k-x[m]) =\hat{p}_k[m], \qquad (k,m) \in {\cal S}
\label{estcdf}
\end{equation}
where $x[m]$ represents the input signal amplitude associated with the estimated probability $\hat{p}_k[m]$.}
 In practice, $x[m]$ is not known. However 
 once the signal parameters are estimated, an estimate {\color{black}$\hat{x}[m] =S[m]^T \hat{\Theta}$}
 of $x[m]$ is available and can be substituted in (\ref{estcdf}), as follows:
\begin{equation}
	\hat{F}_{\eta}(T_k-\hat{x}[m]) =\hat{p}_k[m]. 
\label{estcdf2}
\end{equation} 
In addition, normalization by the estimated standard deviation $\hat{\sigma}$, 
provides an estimate of the CDF of the normalized random variable $\overline{\eta} = \nicefrac{\eta}{\sigma}$.
\begin{equation}
		\hat{F}_{{\eta}}(T_k-\hat{x}[m])=\hat{F}_{\overline{\eta}}\left(\frac{T_k-\hat{x}[m]}{\hat{\sigma}}\right) =\hat{p}_k[m]. 
\label{estcdf3}
\end{equation} 
To validate the initial assumption about the noise CDF,
estimates provided by (\ref{estcdf2}) or (\ref{estcdf3}) can be interpolated and compared
to the assumed noise CDF, e.g. $\Phi(\cdot)$. Then, the corresponding PDF can be estimated by differentiation.
{\color{black} Observe that 
the LSE provides an error sequence obtained as the difference between the estimated signal at the ADC input and the measured signal at the ADC output. However, the histogram of such error samples would not estimate the PDF of the noise at the quantizer input, since the error sequence also contains the error contributions due to quantization. This is not the case with the estimator (\ref{estcdf2}) that is only marginally affected by signal quantization.} 

\section{Simulation Results}
The {\color{black} QBE estimator} was coded in C and simulated on a personal computer using the Monte Carlo approach.
The practical case of estimating the parameters of a sine wave was considered after modeling
the signal as in (\ref{modelsine}).  
Results obtained using the QBE under the assumption of known and unknown signal frequencies and 
known uniformly and non-uniformly distributed transition levels are compared in the following with results obtained using the {\em sine fit} estimation method based on the LSE \cite{Std1241}. 
In all cases the noise standard deviation was assumed unknown, $\lambda$ was set to $0.1155545 \cdots$ and $\epsilon = 0.0011$ was assumed.
As a performance criterion,
the root-mean-square error (RMSE) based on $R$ records of $N$ samples was considered.
This was defined as:
\begin{equation}
	RMSE = \sqrt{e_{DC}^2+\frac{1}{2}e_{AC}^2},
	\label{rmse}
\end{equation}
where $e_{DC}$ and $e_{AC}$ represent the errors in estimating the DC and AC signal components
with respect to the known simulated values. 


\begin{figure}[t!]
\centering
\includegraphics[scale=0.45]{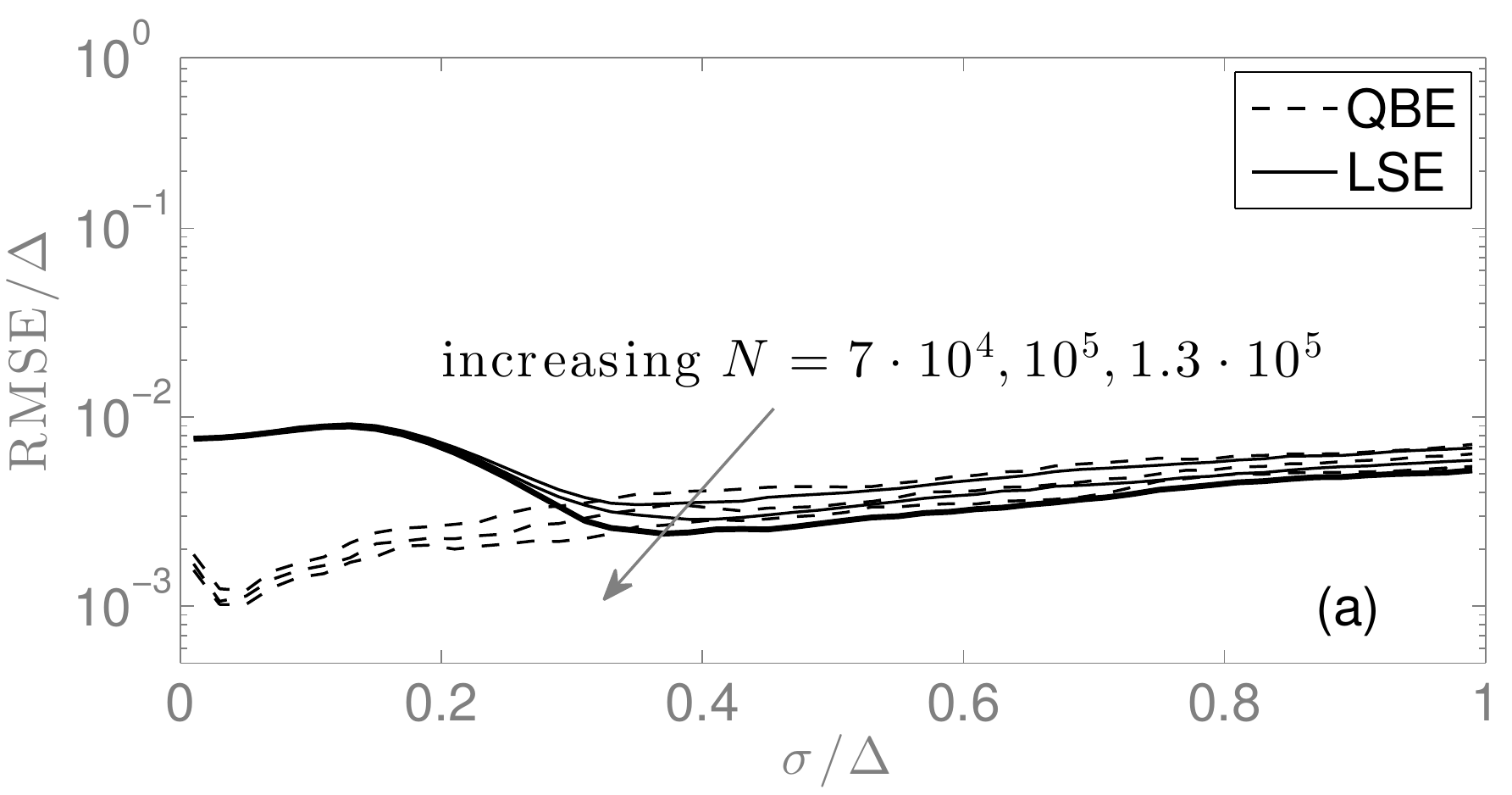}
\includegraphics[scale=0.45]{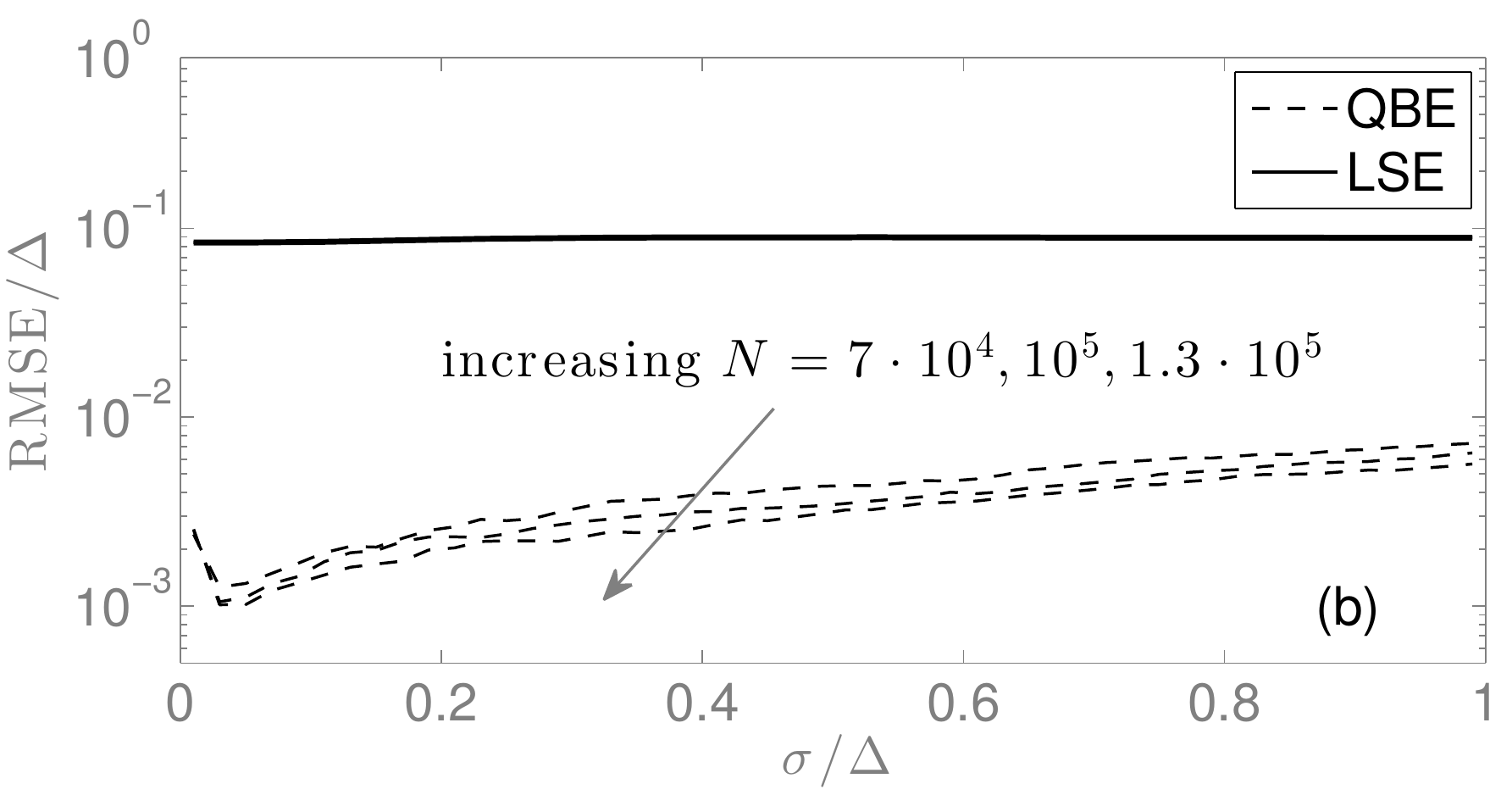}
\includegraphics[scale=0.45]{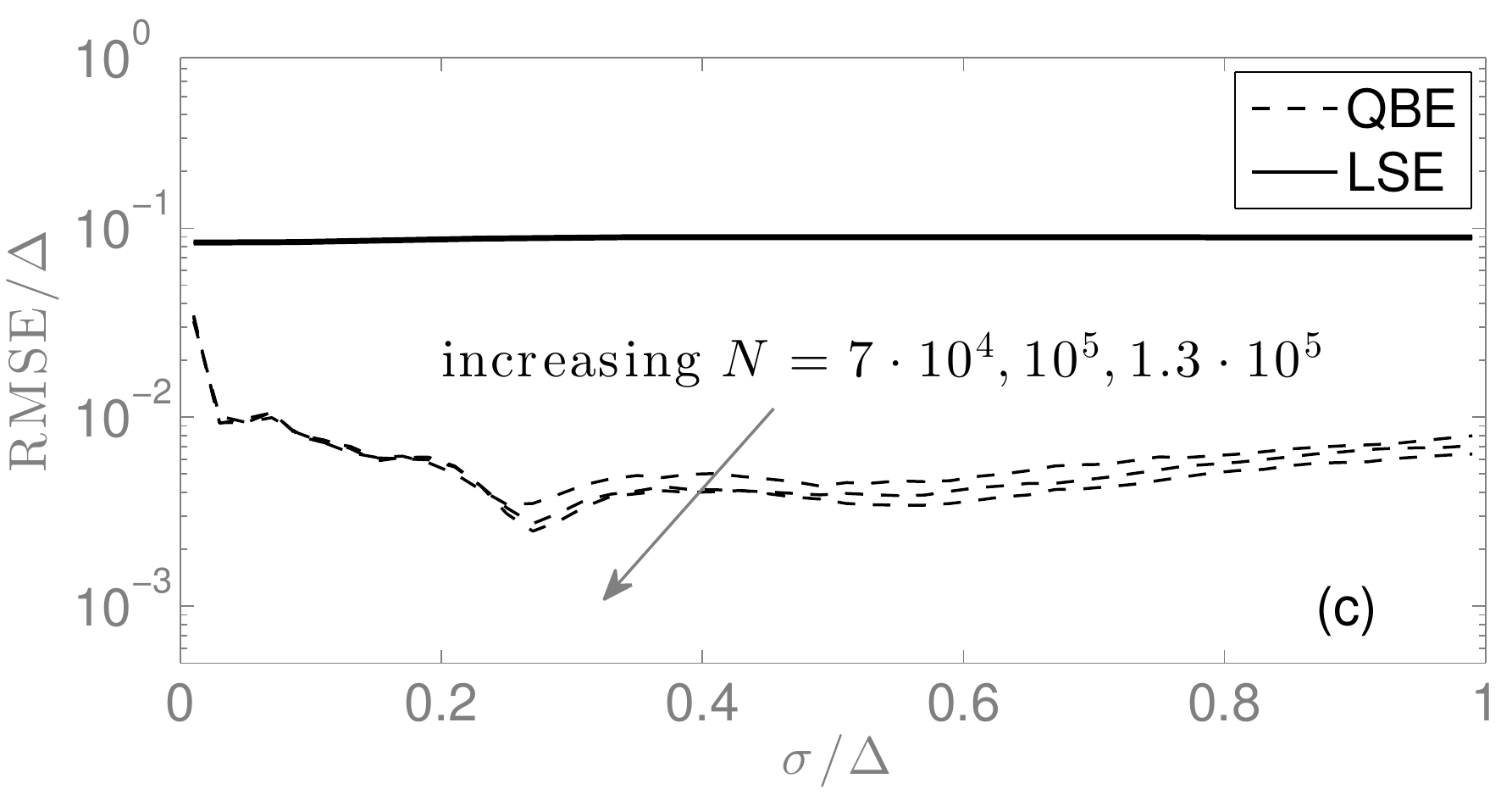}
\caption{Simulation results, known frequency ratio, $\lambda = 0.1155545 \cdots$: 
Root mean-square-error as a function of the 
noise standard deviation (both normalized to $\Delta=20/2^8$)  
{\color{black} when} $N=7\cdot 10^4, 10^5, 1.3\cdot 10^5$,
in the case of the QBE (dashed lines) and LSE (continuous line): (a)
$8$-bit ADC with threshold levels uniformly distributed in the $[-10 V, 10 V]$ input range; 
(b) $8$-bit non-uniform ADC simulated using a resistor ladder with Gaussian distributed resistance and 
maximum absolute INL $ =0.215\Delta$; (c) same as (b) but with additional uncertainty on the values of the transition levels: each transition level is assumed to be known up to a random deviation uniformly 
distributed in the interval $[-0.2 \Delta, 0.2 \Delta]$. \label{simuA}}
\end{figure}

\begin{figure}[t!]
\centering
\includegraphics[scale=0.45]{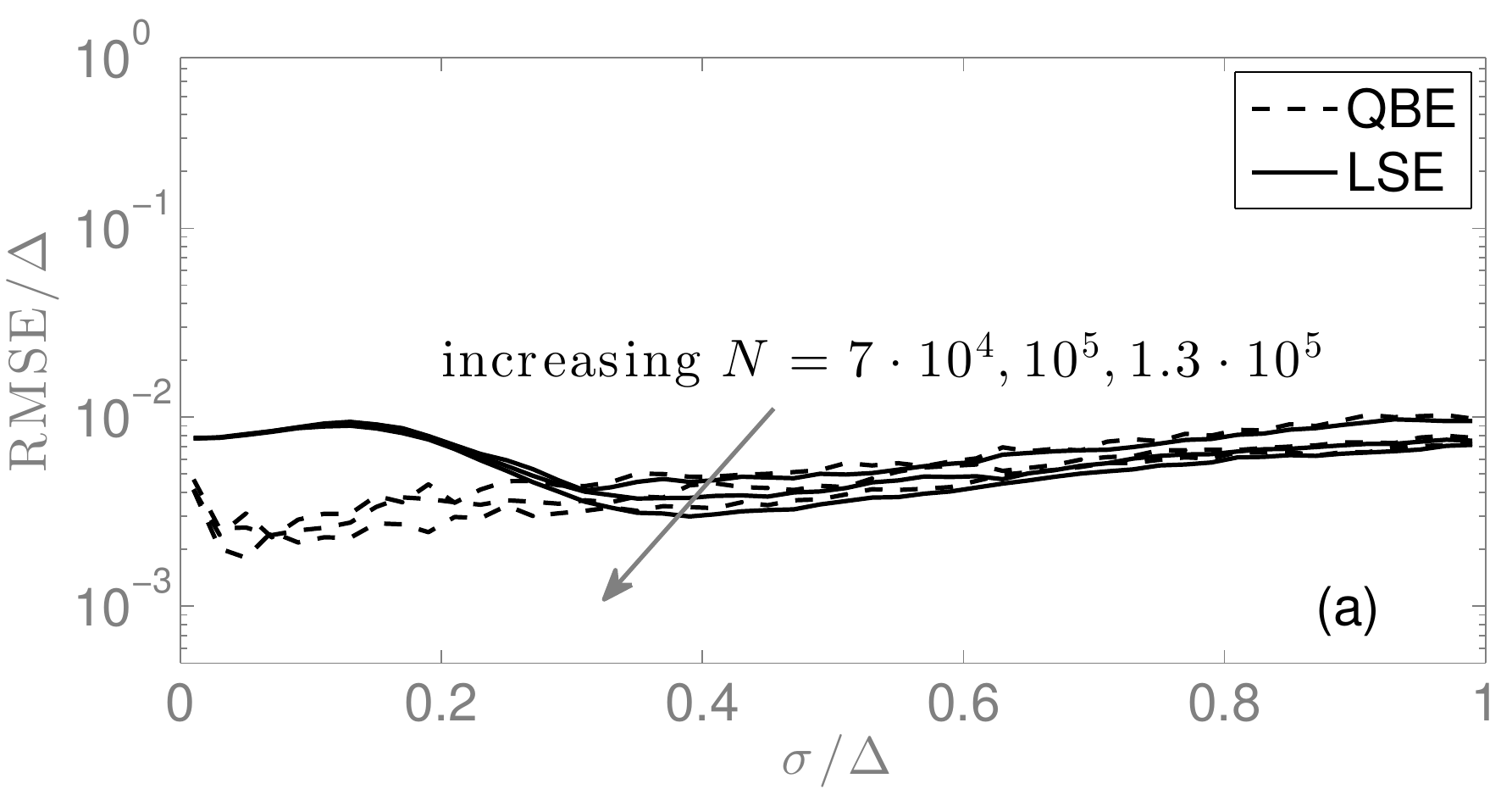}
\includegraphics[scale=0.45]{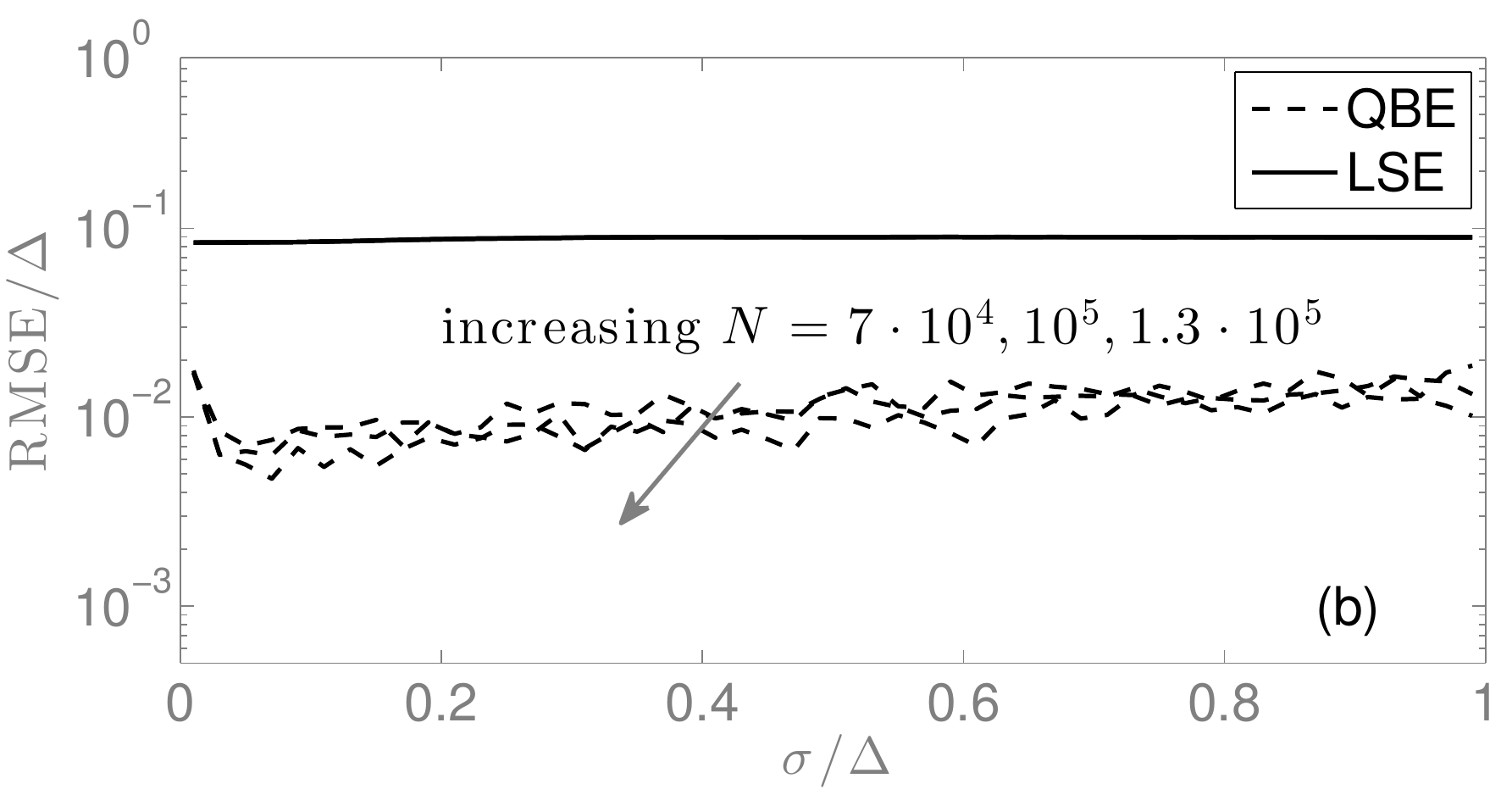}
\caption{Simulation results, unknown frequency ratio, $\lambda = 0.1155545 \cdots$: 
Root-mean-square error as a function of the 
noise standard deviation (both normalized to $\Delta=20/2^8$), and parametrized 
by $N=7\cdot 10^4, 10^5, 1.3\cdot 10^5$,
in the case of the QBE (dashed lines) and LSE (continuous line): (a)
$8$-bit ADC with threshold levels uniformly distributed in the $[-10 V, 10 V]$ input range; 
(b) $8$-bit non-uniform ADC simulated using a resistor ladder with Gaussian distributed resistance and 
maximum absolute INL $= 0.215\Delta$. \label{simuB}}
\end{figure}

\subsection{Known Frequency Ratio $\lambda$\label{simuuA}}
If both the sine wave frequency and the ADC sampling rate are known,
so is the ratio $\frac{T_s}{T}$ and both QBE and the  
$3$-parameter sine fit,  can be applied as described.
Accordingly, simulations were done 
assuming an $8$-bit ADC, $R=100$ records, and
$N=7\cdot 10^4, 10^5, 1.3\cdot 10^5$ samples. The RMSE is graphed 
in Fig.~\ref{simuA} in the case of the QBE (dashed line) and the LSE (solid line),
as a function of the noise standard deviation {\color{black}and for various values of $N$}. Both axes are normalized to the 
quantization step $\Delta=20/2^8$.
While data in Fig.~\ref{simuA}(a) refer to the case of a uniform ADC,
graphs in   Fig.~\ref{simuA}(b) are associated {\color{black}with} a non-uniform 
ADC, based on a resistor ladder \cite{Maloberti}. 
Distribution of resistance values following a Gaussian distribution resulted in a maximum absolute INL of ${\color{black} 0.215\Delta} $. 
It can be observed that:
\begin{itemize}
\item for a given value of $\sigma$, when $N$ increases, the RMSE shows an overall decrease, as expected;
\item when $\sigma/\Delta$ is small, the RMSE associated {\color{black}with} the LSE is dominated 
by the estimator bias, rather than by its variance: in fact, by increasing $N$, 
the corresponding RMSE does not change in the left part of the graph (solid lines). This means that 
the contribution of the estimation variance to the RMSE, that depends on $N$, is negligible with respect to
the bias and that the bias does not vanish when $N$ increases, as expected \cite{CarboneSchoukens}; 
\item the RMSE associated {\color{black}with} QBE is largely independent {\color{black}of} the ADC being uniform or non-uniform since it only uses information about threshold levels, irrespective of their distribution over the input range. Conversely, 
since the LSE processes code values, departure from uniformity in the distribution of the transition levels results
in an overall worse performance, as shown by comparing data in Fig.~\ref{simuA}(a) to data in Fig.~\ref{simuA}(b).
This latter figure shows that the RMSE in the case of the LSE (solid lines), is dominated by estimation 
bias rather than estimation variance, since all curves collapse, irrespective of the 
 number of processed samples.
\end{itemize}
  
The QBE relies on the knowledge of the ADC transition levels that are known, in practice, only through measurement results affected by uncertainty. 
To show the robustness of the QBE with respect to this aspect, 
a simulation was done by assuming the transition levels known
up to a random deviation uniformly distributed in the interval $[-0.2 \Delta, 0.2\Delta]$.  The same data used 
to produce graphs in Fig.~\ref{simuA}(b) was used under the same simulated conditions. Results are shown in Fig.~\ref{simuA}(c), which displays an increase of the RMSE evident for small values of $\sigma$, but also that the QBE still outperforms the LSE when INL affects the quantizer.
  
\subsection{Unknown Frequency Ratio $\lambda$}
If either or both signal frequency and sampling 
rate are unknown, so is the ratio
$\frac{T_s}{T}$.
In this case, the procedure described in subsection~\ref{ukn} applies. After a rough estimation of
$\lambda$ based on the discrete Fourier transform of the simulated data, the estimation 
procedure is applied iteratively to find the 
minimum of the experimental mean-square-error, defined as:
\begin{equation}
	\mbox{MSE}_{\mbox{exp}} = {\frac{1}{N}\sum_{n=0}^{N-1} {\left( \overline{S}[n]^T\hat{\Theta}-x_q[n]\right)^2}},
\label{mseiter}
\end{equation}
where {\color{black}$\overline{S}[n]^T\hat{\Theta}$} is 
the estimated input signal at time $n$ and $x_q[n]$ the known quantized value.
Observe that (\ref{mseiter}) and (\ref{rmse}) consider different errors: while in 
(\ref{mseiter}) the error is defined with respect to the {\em measured} values $x_q[{\color{black}n}]$, in (\ref{rmse})
the error is defined with respect to the known {\em simulated} values.

An estimate of (\ref{mseiter}) is found by 
minimizing  $\mbox{MSE}_{\mbox{exp}}$ over the set of possible values of 
$\Theta=\left[\frac{\theta_0}{\theta_3}\; \frac{\theta_1}{\theta_3}\; \frac{\theta_2}{\theta_3}\; \frac{-1}{\theta_3}\right]$ through the golden section search algorithm \cite{golden}.
Here $\theta_0, \theta_1,$ and $\theta_2$ represent the {\em sine}, {\em cosine} and {\em dc} components, respectively, as defined in $(\ref{modelsine})$, while $\theta_3$ represents 
the unknown noise standard deviation. 
The resulting RMSE is shown in Fig.~\ref{simuB}(a) and (b) in the case
of a uniform and non-uniform ADC, respectively. 
Same simulated conditions as in subsection \ref{simuuA} were applied in this case, but with $R=30$.
Although the frequency was not assumed as being known in advance, 
results shown in Fig.~\ref{simuB} are comparable
to those graphed in Fig.~\ref{simuA}.

\begin{figure}[b!]
\begin{center}
\includegraphics[scale=0.35]{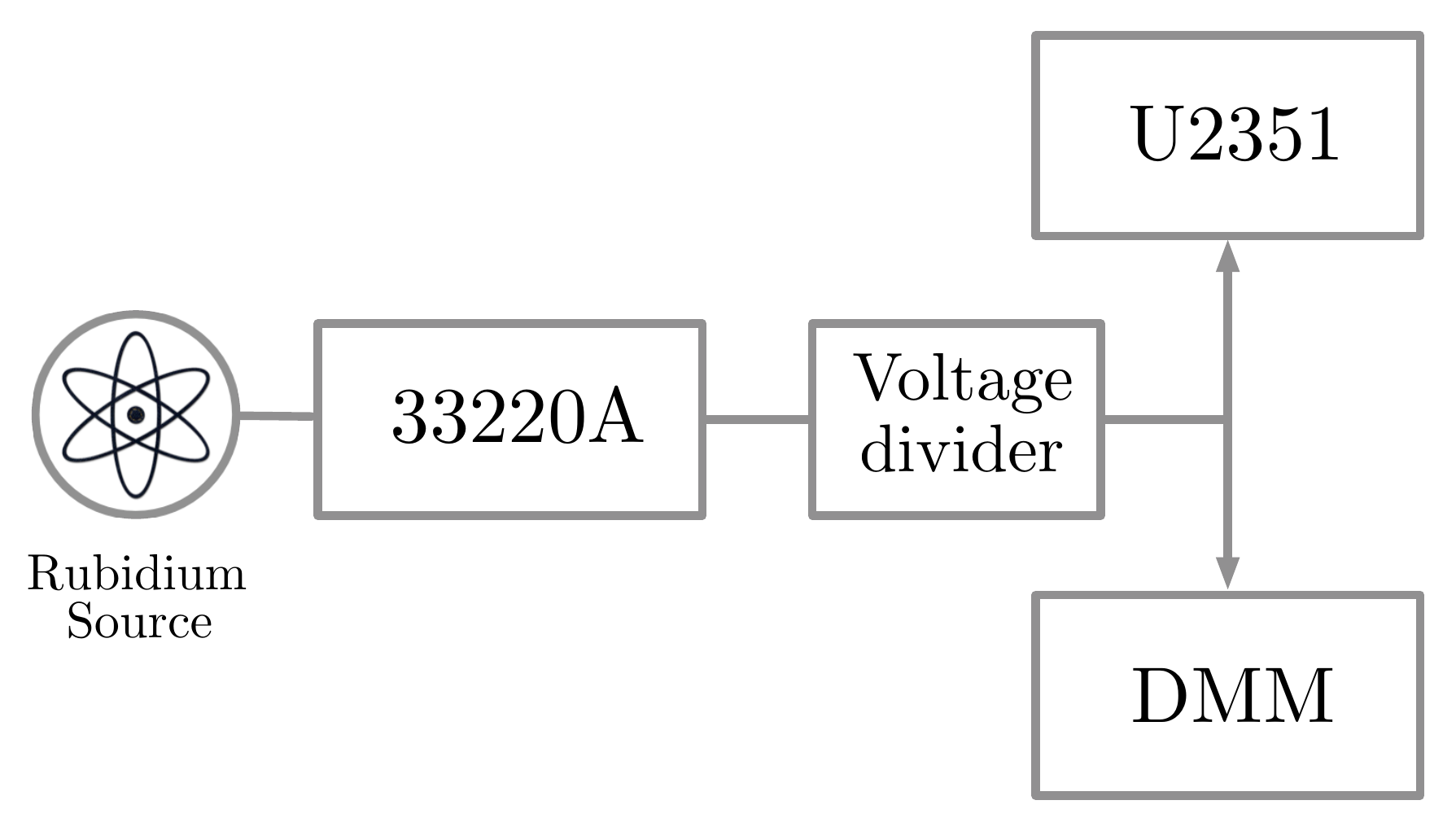}
\caption{Measurement setup used for the experiments. The Rubidium frequency standard Stanford Research System PRS10 is used to provide a stable clock to the waveform synthesizer Agilent 33220A. This is used to generate the test signals that are input to the $16$-bit DAQ Keysight U2351. A resistor-based voltage divider is used to reduce the range of the generated signals. 
A $6$-$\nicefrac{1}{2}$ digit multimeter (DMM, Keithley 8845A) is used as the reference instrument, measuring both DC and AC signals. A personal computer controls the measurement chain. \label{meascheme}}
\end{center}
\end{figure}
\section{Experimental Results}
The measurement setup shown in Fig.~\ref{meascheme} was used to first measure the ADC transition levels and
then to perform measurements {\color{black} to estimate the sine wave parameters}, the noise standard deviation and its CDF. 
It included a rubidium frequency standard used to control a waveform synthesizer. The generated signal was acquired both by a USB connected 16-bit DAQ (U2351) by Keysight Technologies and by a $6\nicefrac{1}{2}$ digit DMM, whose results were taken as reference values. 
The voltage divider was used to reduce the range of values generated 
by the waveform synthesizer by a factor approximately equal to $30$. The exact attenuation factor was not needed 
because all voltages were also measured by the reference instrument.

\begin{figure}[tp!]
\begin{center}
\includegraphics[scale=0.45]{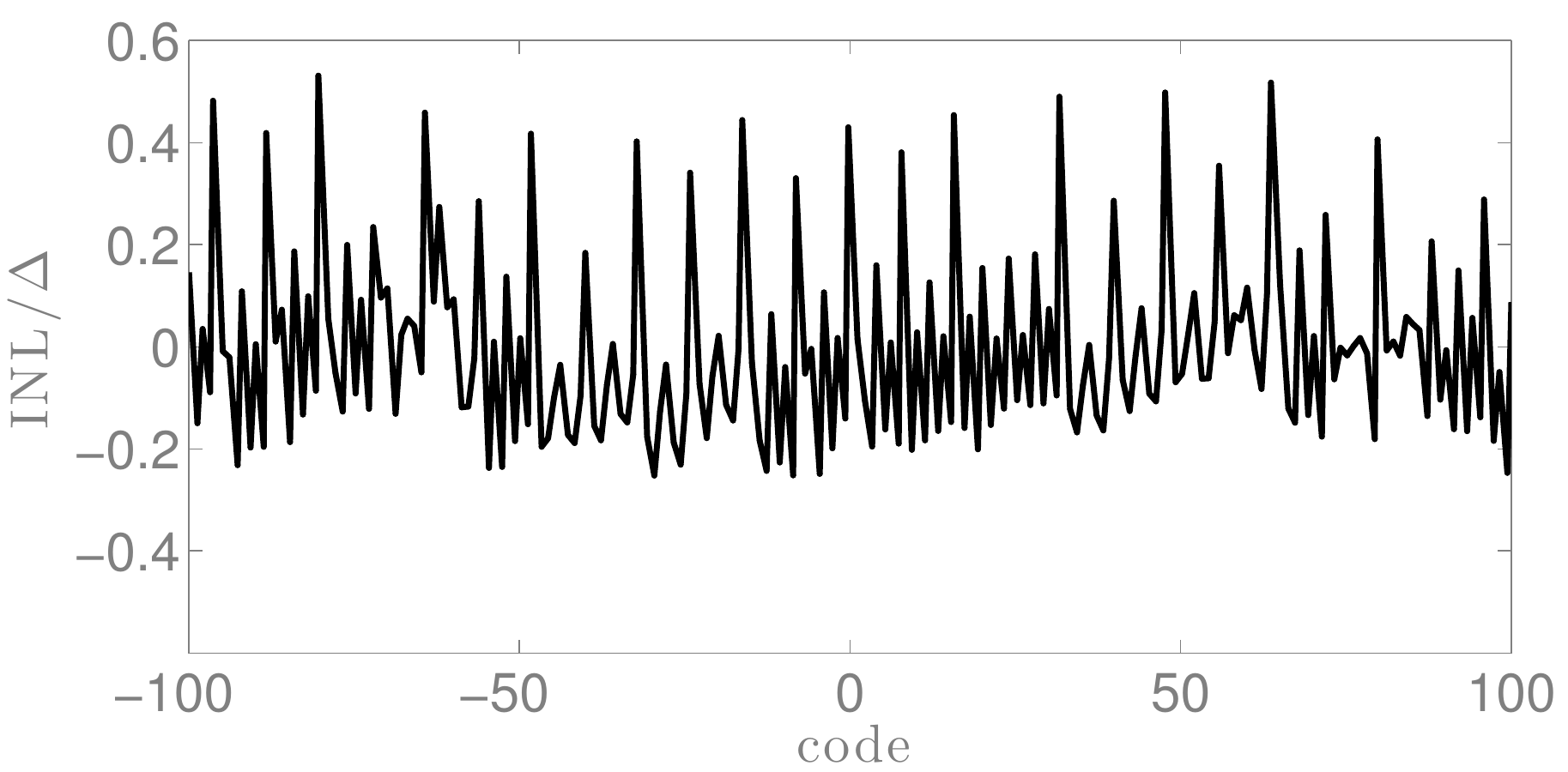}
\caption{INL normalized to the quantization step $\Delta=\nicefrac{20}{2^{16}}$ and measured through the measurement setup shown in Fig.~\ref{meascheme}, after removal of gain and offset errors. \label{measINL}}
\end{center}
\end{figure}

\begin{figure}[t!]
\begin{center}
\includegraphics[scale=0.45]{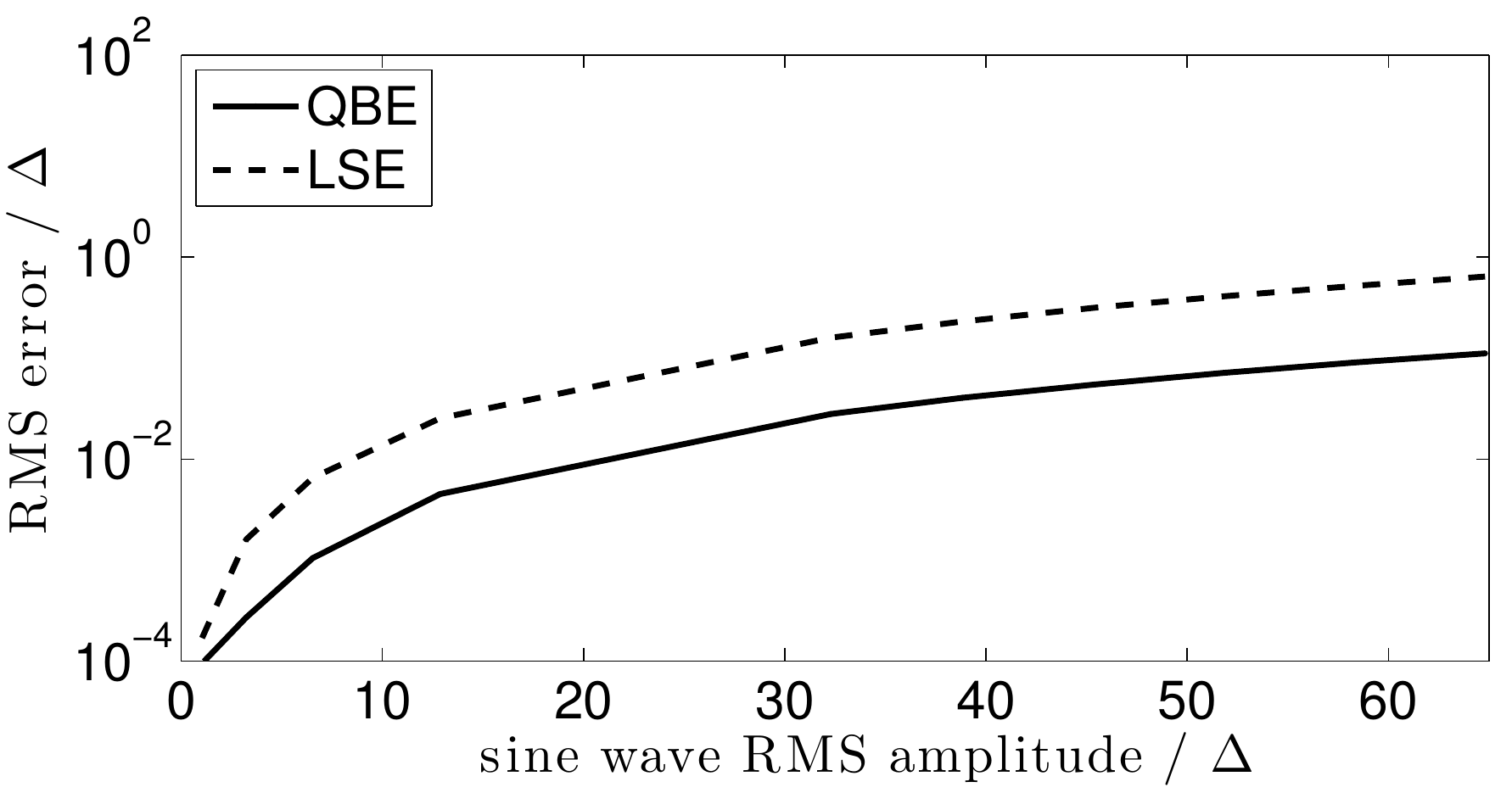}
\caption{Experimental data: performance comparison between the QBE and the LSE; RMSE in the estimation of the DC and AC values of a sine wave generated and quantized using the setup shown in Fig.~\ref{meascheme}. 
Both axes are normalized to the quantization step $\Delta=\nicefrac{20}{2^{16}}$ and
both the QBE and LSE estimates are based on $N=1.5\cdot 10^5$ samples. The ratio between sine wave frequency and sampling rate was estimated by finding the minimum of the square error cost function, using the iterative golden section search algorithm.
\label{rmserror}}
\end{center}
\end{figure}

The DAQ transition levels were first measured in the code interval $[-100,100]$, using a software 
implementation of the servo-loop technique \cite{Std1241}. 
Measured values were fitted using linear interpolation to remove gain and offset errors and to obtain the 
integral nonlinearity shown in Fig.~\ref{measINL}, after normalization to the DAQ quantization step $\Delta=\nicefrac{20}{2^{16}}\;$ V. 
The synthesizer was programmed to generate a sine wave  with nominal 
frequency $500$ Hz, sampled by the DAQ at $500$ kSample/s in the 
$[-10, 10]$ V input range.   
The procedure processed $10$ sine wave amplitudes in the range $[1.042 \Delta, 64.803 \Delta]$. 
For each amplitude{\color{black},} $N=1.5\cdot 10^5$ samples were collected and processed by QBE and LSE iteratively.
The DMM was programmed to measure each time the AC and DC signal components in average mode.   
These values were taken as the true values of the sine wave parameters and (\ref{rmse}) {\color{black} was} then applied
to evaluate the RMSE. 
This figure is graphed in Fig.~\ref{rmserror} for both estimator, as a function of the AC signal component.
Data show that QBE outperforms the LSE.
In fact, contrary to QBE, the solution provided by LSE ignores the effect of INL on measured data.

QBE also provided an estimate of the noise standard deviation and of the noise CDF.
Estimates of the noise standard deviation were obtained 
for each one of the $10$ datasets collected by varying the sine wave amplitude.
The estimated mean value and standard deviation of these $10$ estimates were $0.800\Delta$ and 
about $0.004 \Delta$, respectively. Thus, stable and repeatable results were obtained.

\begin{figure}[t!]
\begin{center}
\includegraphics[scale=0.45]{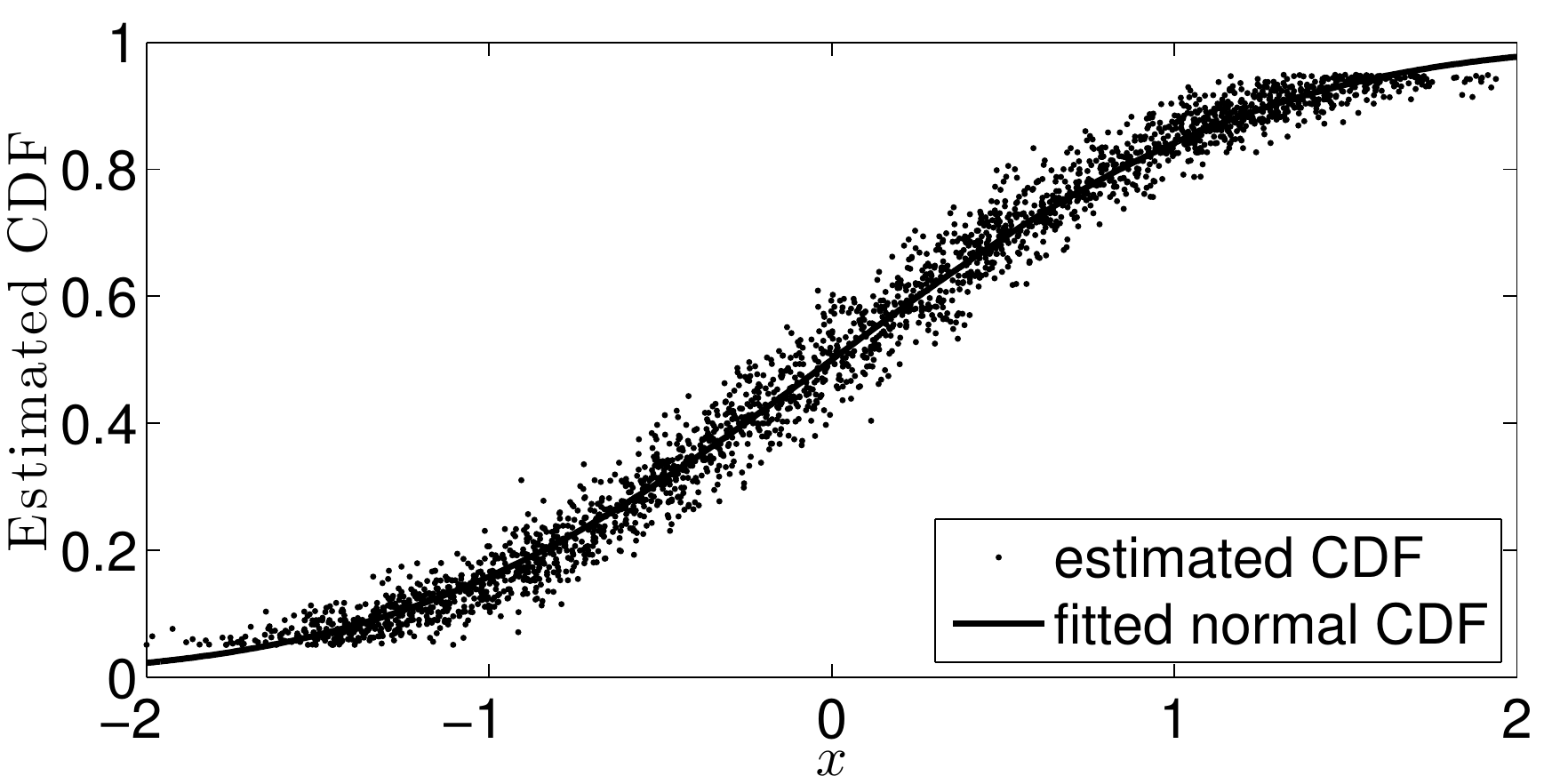}
\caption{Experimental results obtained using the measurement setup shown in Fig.~\ref{meascheme}.
Pointwise estimation of the normalized input noise CDF obtained by (\ref{estcdf3}) (dots)
and plot of the CDF of a standard Gaussian random variable (solid line). The input sine wave had a measured RMS amplitude of
$3.93\cdot 10^{-3}$ V and a measured DC component of $7.9\cdot 10^{-5}$ V. 
 \label{expcdf}}
\end{center}
\end{figure}

{\color{black}The} estimated CDF associated {\color{black}with} the noise sequence normalized to the standard deviation is plotted in Fig.~\ref{expcdf} along with a fitted CDF of the standard Gaussian CDF. The good match between 
the points and the fitted curve validates both the assumption {\color{black}of} the noise distribution and the correctness of the adopted approach. 
Notice that the choice of the 
guard intervals $[0, 0.05]$ and $[0.95,1]$ 
used for discarding selected data, 
resulted in an  
estimate of the noise CDF that is truncated in the bottom and upper parts of the graph.

\section{Conclusion}
Signal quantization is customarily performed in numerical 
instrumentation. 
Even if research activities continuously 
provide new ADC architectures exhibiting 
increasing performance, devices
are not characterized by uniformly distributed transition levels. 
If not compensated,
this non-ideal behavior results in biases and distortions in the estimated quantities. 

In this paper, we proposed a new estimator that  {\color{black}uses} information about 
the values of the ADC transition levels to improve the performance of conventional estimators.
The estimator is based on the knowledge of the ADC transition levels, so that an initial calibration 
phase is necessary before actual parameter estimation.
Theoretical, simulated and experimental results show that the proposed technique outperforms 
typically used estimators such as the sine fit, 
both when the input signal frequency {\color{black}is} known and unknown. As an additional benefit, 
also the ADC input noise standard deviation and the noise 
cumulative distribution function are estimated. 
Finally, 
while results are presented mainly in the context of the estimation of the parameters of a sine wave, this procedure can be applied whenever a bounded periodic signal is acquired and processed.

\section*{Acknowledgement}
This work was supported in part by the Fund for Scientific Research (FWO-Vlaanderen), by the Flemish Government (Methusalem), the Belgian Government through the Inter university Poles of Attraction (IAP VII) Program, and by the ERC advanced grant SNLSID, under contract 320378.
\balance


\end{document}